\documentclass[prl,twocolumn,preprintnumbers,amsmath,amssymb,floatfix]{revtex4}
\usepackage{graphicx}
\usepackage{bm}
\usepackage{color}

\def\nablab{{\bm \nabla}}

\def\J{\mathcal{J}}
\def\O{\mathcal{O}}

\def\ExB{{\bm E}\times{\bm B}}

\def\figures{.}

\definecolor{gray}{rgb}{0.5,0.5,0.5}
\definecolor{dred}{rgb}{0.5,0.0,0.0}
\definecolor{dgreen}{rgb}{0.0,0.5,0.0}
\definecolor{dblue}{rgb}{0.0,0.0,0.5}
\definecolor{violet}{rgb}{0.7,0.0,0.5}

\definecolor{lred}{rgb}{1.0,0.5,0.5}
\definecolor{lgreen}{rgb}{0.5,1.0,0.5}
\definecolor{lblue}{rgb}{0.5,0.5,1.0}

\begin{document}

\preprint{}

\title{\vspace*{-0.6cm} Energy-selective confinement of fusion-born alpha particles \\ during internal relaxations in a tokamak plasma\vspace*{-0.2cm}} 

\author{A.~Bierwage$^{1}$\footnote{Electronic address: {\tt bierwage.andreas@qst.go.jp}}, K.~Shinohara$^{2,1}$, Ye.O.~Kazakov$^{3}$, V.~Kiptily$^{4}$, Ph.~Lauber$^{5}$, M.~Nocente$^{6,7}$, \v{Z}.~\v{S}tancar$^{8,4}$, S.~Sumida$^{1}$,  M.~Yagi$^{1}$, J.~Garcia$^{9}$, S.~Ide$^{1}$ and JET Contributors\footnote{See the author list of `Overview of JET results for optimising ITER operation', J. Mailloux {\it et al.}, {\it Nucl.\ Fusion} {\bf 62} (2022) ??????.}}

\affiliation{$^{1}$ QST, Naka and Rokkasho Fusion Institutes, Japan \\
$^{2}$ \mbox{Department of Complexity Science and Engineering, The University of Tokyo, Kashiwa, Chiba, Japan} \\
$^{3}$ \mbox{Laboratory for Plasma Physics, LPP-ERM/KMS, Partner in the Trilateral Euregio Cluster (TEC), Brussels, Belgium} \\
$^{4}$ \mbox{Culham Centre for Fusion Energy of UKAEA, Culham Science Centre, Abingdon, United Kingdom} \\ 
$^{5}$ Max-Planck-Institut f\"{u}r Plasmaphysik, Garching, Germany \\
$^{6}$ \mbox{Dipartimento di Fisica `G. Occhialini’, Universit\`{a} di Milano-Bicocca, Milano, Italy} \\
$^{7}$ \mbox{Institute for Plasma Science and Technology, National Research Council, Milan, Italy} \\
$^{8}$ Jo\v{z}ef Stefan Institute, Ljubljana, Slovenia \\
$^{9}$ CEA, IRFM, 13108, Saint-Paul-lez-Durance, France}

\date{\today}

\begin{abstract}\vspace*{-0.2cm}
Long-pulse operation of a self-sustained fusion reactor using toroidal magnetic containment requires control over the content of alpha particles produced by D-T fusion reactions. On the one hand, MeV-class alpha particles must stay confined to heat the plasma. On the other hand, decelerated helium ash must be expelled before diluting the fusion fuel. Our kinetic-magnetohydrodynamic hybrid simulations of a large tokamak plasma confirm the existence of a parameter window where such energy-selective confinement can be accomplished by exploiting internal relaxation events known as `sawtooth crashes'. The physical picture --- consisting of a synergy between magnetic geometry, optimal crash duration and rapid particle motion --- is completed by clarifying the role played by magnetic drifts. Besides causing asymmetry between co- and counter-going particle populations, magnetic drifts determine the size of the confinement window by dictating where and how much `reconnection' occurs in particle orbit topology.
\end{abstract}


\maketitle

\thispagestyle{empty}
\everypar{\looseness=-1} 

\section{Introduction}

Laboratories around the world have intensified R\&D activities for experimental reactors that should demonstrate the practical feasibility of extracting useful energy from controlled nuclear fusion. Spearheaded by ITER \cite{iter}, the tokamak concept is the present mainstream approach to magnetically confined fusion (MCF). While their use as a power plant still awaits breakthroughs, the accumulated scientific evidence suggests that tokamaks can produce a `burning plasma', where fusion reactions are self-sustained for times much longer than the confinement times of thermal energy and charged particles.

\begin{figure}
[tbp]
\centering\vspace{-0.35cm}
\includegraphics[width=0.45\textwidth]{\figures/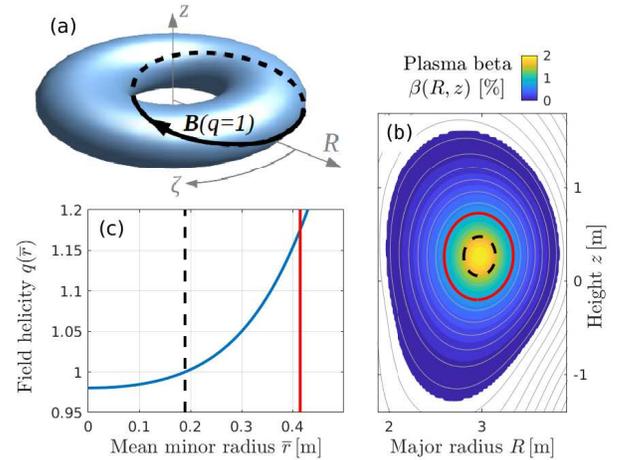}
\caption{(a) Toroidal geometry of a tokamak in cylinder coordinates $(R,z,\zeta)$. The black curve represents a magnetic field line with helicity $q = 1$. (b) Shape of the plasma cross-section in a poloidal $(R,z)$ plane in our working example based on the JET tokamak. The gray contours are magnetic flux surfaces. The color contours show the plasma beta, which measures the ratio of thermal to magnetic pressure as $\beta = 2\mu_0 P/B_0^2$ with $\mu_0 = 4\pi\times 10^{-7}\,{\rm H/m}$. (c) Central portion of a field helicity profile $q(\overline{r})$ that facilitates energy-selective alpha particle confinement. The dashed black line marks the $q=1$ surface. The red line is the boundary of a reduced simulation domain.}\vspace{-0.25cm}
\label{fig:01_equil}%
\end{figure}

Tokamaks use a strong magnetic field to confine hydrogen isotope plasmas with high temperatures ($\sim 10\,{\rm keV}$) in a toroidal volume as sketched in Fig.~\ref{fig:01_equil}(a). The helically wound ${\bm B}$ field consists of a dominant toroidal component ${\bm B}_{\rm tor}$ that is provided by external coils and a weaker poloidal component ${\bm B}_{\rm pol}$ that is induced by electric currents carried by the plasma itself. For a toroidal surface with long circumference $2\pi R$ and short circumference $2\pi\overline{r}$, the mean helical pitch of the magnetic field vector ${\bm B}$ is given by $q \approx \overline{r} B_{\rm tor}/(R B_{\rm pol})$, where $R$ is the major radius of the torus and $\overline{r}$ the mean minor radial distance from the center of the plasma. In preparation for a later generalization, we call $q$ the `field helicity' \cite{WhiteTokBook3}.

The field helicity profile $q(\overline{r}) \propto 1/I_{\rm tor}(\overline{r})$ varies across the plasma radius, in inverse proportion to the plasma's electric current profile $I_{\rm tor}(\overline{r})$. Each toroidal surface where the field helicity has a rational value $q = m/n$ represents a geometric resonance. Resonances with small integers $m$ and $n$ can facilitate macroscopic long-lived plasma distortions and instabilities. Most notably, when the plasma current distribution reaches a certain threshold, so that the field helicity drops below unity ($q < 1$) somewhere in the plasma, a self-organization process sets in that prevents further steepening of the current density profile \cite{Shafranov70}. This results in a quasi-steady state, which future fusion reactor experiments like ITER \cite{iter} will exploit.

\begin{figure}
[tbp]
\centering\vspace{-0.25cm}
\includegraphics[width=0.45\textwidth]{\figures/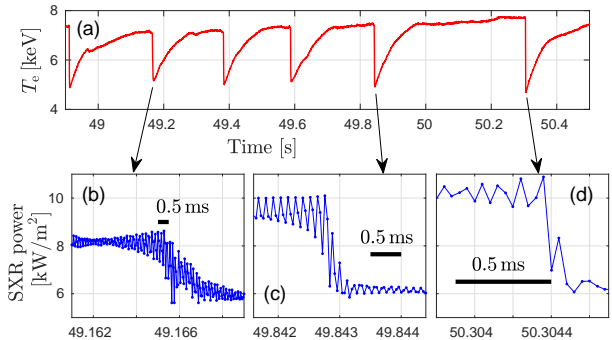}
\caption{Examples of sawtooth crashes in JET pulse 95679 \protect\cite{Nocente20,Kazakov21} as seen by electron temperature fluctuation measurements in the plasma core using (a) electron cyclotron emission (ECE, red) and (b-d) soft x-ray (SXR, blue) diagnostics.\vspace{-0.15cm}}
\label{fig:02_jet_saw_raw}%
\end{figure}

This self-organization process can be pictured as follows. $q = 1$ means that magnetic field lines close on themselves after one poloidal and one toroidal turn as illustrated in Fig.~\ref{fig:01_equil}(a). In our example, which has the dimension of the Joint European Torus (JET), the condition $q < 1$ is satisfied within a radius of about $\overline{r} \approx 0.2\,{\rm m}$ as indicated by the dashed lines in panels (b) and (c) of Fig.~\ref{fig:01_equil}. The portion of the plasma located within the $q = 1$ surface can be easily displaced by the destabilizing forces associated with gradients in the current density. The resulting perturbation, which is known as `internal kink' mode, has the form of a `tilted torus within a torus'. In other words, the kinked $q=1$ torus is resonant with the toroidal geometry of the tokamak as a whole. Together with mechanisms facilitating magnetic reconnection \cite{Kadomtsev75}, the formation of a region with $q < 1$ gives rise to quasi-periodic relaxation events that can be observed in the form of `sawtooth oscillations' in time traces of the central electron temperature $T_{\rm e}$ as in Fig.~\ref{fig:02_jet_saw_raw}. These data were acquired during a JET pulse, where the 3-ion radio-frequency (RF) heating scheme was applied to a mixed D-$^3$He plasma, successfully generating and confining fusion-born alpha particles \cite{Nocente20, Kazakov21, Kiptily21}.

Benign sawtooth activity is considered to have beneficial effects in fusion-oriented applications. Besides helping to keep the core plasma near a well-defined state, this mixing process was found to prevent excessive accumulation of heavy-ion impurities that would cause radiative cooling \cite{Nave03}. By the same token, it has repeatedly been proposed to use sawteeth for the expulsion of helium ash (henceforth called `slow alphas') from the core of a deuterium-tritium (D-T) fusion reactor.

There is however a caveat: It is not desirable to have sawteeth that flatten the profiles of all particles. In particular, energetic $^4_2{\rm He}^{2+}$ ions (`fast alphas') should ideally be left unperturbed, since they provide the heating power in a self-sustained `burning' fusion plasma.

After decades of research --- especially during the 1990s when the Tokamak Fusion Test Reactor (TFTR) and JET operated with D-T plasmas \cite{Strachan97, JET_DT99} --- this conundrum of needing to confine fast alphas while expelling slowed-down helium ash is still being actively explored, with or without sawteeth \cite{White21a, White21b}.

Until recently, the existing evidence indicated that the majority of particles, including fast alphas, will undergo mixing during a sawtooth crash, so that their density profiles are flattened in the relaxation domain (unless, of course, the profile had already been broad before the crash \cite{Marcus94}). In a 2014 review \cite{Gorelenkov14}, it was concluded that ``the effects of kink modes on fast ions seem to be understood.'' This was followed by the construction of computationally efficient reduced models, which are being used in integrated transport simulations to make quantitative predictions for experiments such as ITER, assuming that all sawtooth crashes have the same effect \cite{Mirnov18}.

It turns out, however, that the $q$ profile can have a significant influence on the transport of alpha particles. Using kinetic-magnetohydrodynamic hybrid simulations, we confirm predictions that there exists a parameter window where MeV-class alphas can sustain a sharply peaked density profile even inside the sawtooth mixing radius, whereas the majority of partially slowed-down alphas with energies of a few $100\,{\rm keV}$ or less are strongly mixed. This is explained in terms of a synergistic effect: During a type of sawtooth where the magnetic field helicity always remains close to unity ($q \sim 1$), so that the particle orbit topology is sensitive to drifts associated with magnetic gradients ($\nablab B$ and curvature), the typical time scale of a sawtooth crash turns out to be just right for it to act differently on slow and fast alphas.

The fact that sufficiently energetic charged particles can decouple from the dynamics of the bulk of a magnetized plasma has been known since the early years of MCF research (see \cite{Porcelli91} for a review). The energy threshold above which ions decouple from the internal kink mode has been estimated theoretically by Kolesnichenko \& Yakovenko \cite{Kolesnichenko96} and confirmed experimentally by Muscatello {\it et al}.\ \cite{Muscatello12}. Using a heuristic model of a sawtooth crash, Jaulmes {\it et al}.\ \cite{Jaulmes14} simulated the redistribution of alpha particles in a JET-like configuration, showing that the threshold energy can be expected to lie at a few $100\,{\rm keV}$ for trapped, and above $1\,{\rm MeV}$ for passing alpha orbits when the parameter $|1 - q|$ is sufficiently small.

Besides confirming the predictions of the heuristic models, we complete the physical picture by clarifying the role of `reconnection' in orbit topology. In fact, some of the approximations used in \cite{Kolesnichenko96} break down when the magnitude of the parameter $|1 - q|$ drops to the level of a few percent, where the resonance condition between alpha particles and the kink becomes sensitive to magnetic drifts. We show that this is precisely the regime, where the majority of fast alphas can decouple from the kink and the looked-for energy-selective confinement can be realized for nearly all pitch angles. The intrinsic preference for counter-passing particles to remain better confined than co-passing ones is also explained and discussed.

\vspace{-0.1cm}
\section{Results}\vspace{-0.25cm}

\subsection{Numerical simulation of a reconnecting internal kink}

Our simulations are performed using a so-called hybrid code \cite{Todo98,Todo05}, which solves visco-resistive magnetohydrodynamic (MHD) equations for the bulk plasma and kinetic equations for the fast ion minority species, whose inertia is assumed to be negligible compared to that of the bulk plasma. The plasma size and field strength are based on JET. The magnetic axis has a major radius of $R_0 = 3\,{\rm m}$ and a field strength of $B_0 = 3.7\, {\rm T}$. The plasma current is $I_{\rm p} = 2.5\,{\rm MA}$. The exact plasma composition is irrelevant in our single-fluid MHD model; here, the chosen bulk ion density and effective particle mass yield a central Alfv\'{e}n speed of $v_{\rm A0} \approx 8\times 10^6\,{\rm m/s}$. (For pure deuterium, this corresponds to a central density of about $5.3\times 10^{19}\,{\rm m}^{-3}$.) Figure~\ref{fig:01_equil}(a) shows the plasma torus schematically in right-handed cylinder coordinates $(R,z,\zeta)$. Panel (b) shows the plasma cross-section in the poloidal $(R,z)$ plane, where the toroidal angle $\zeta$, magnetic field vector ${\bm B}$ and plasma current are all pointing out of the plane. Panel (c) shows the assumed profile of the field helicity $q$, which we treat as a free parameter.

For the purpose of this proof-of-principle study, it is not necessary to simulate the reconnection process in all its multi-scale detail, which is extremely challenging even with modern supercomputers \cite{KumarAPS21}. Considering the type of sawtooth that is associated with an internal kink instability and proceeds in a fashion similar to that envisioned by Kadomtsev \cite{Kadomtsev75} and Wesson \cite{Wesson86}, it suffices here to simulate the associated global changes in the magnetic topology and the global electric drift. We require that the plasma is mixed in a similar volume and on a similar time scale of a few $100\,\mu{\rm s}$ as in experiments (Fig.~\ref{fig:02_jet_saw_raw}). By matching the crash time scale and the size of the relaxing domain, we can match the speed of the displacement.

The size of the relaxing domain was inferred from electron temperature measurements like those in Fig.~\ref{fig:02_jet_saw_raw}, which determine the approximate location of the $q = 1$ surface that is indicated by dashed lines in Fig.~\ref{fig:01_equil}. The relaxation region is relatively small, so that it suffices in many of our simulations to cover only the inner 25\% of the plasmas's magnetic flux space. In that case, an artificial non-slip boundary is placed along the red line in Fig.~\ref{fig:01_equil}, at about twice the sawtooth mixing radius. Inside that region we choose the $q$ profile to be close to unity and relatively flat, which yields a configuration that is both numerically tractable and practically relevant for long-pulse scenarios in future ITER experiments.

\begin{figure}
[tbp]
\centering\vspace{-0.25cm}
\includegraphics[width=0.45\textwidth]{\figures/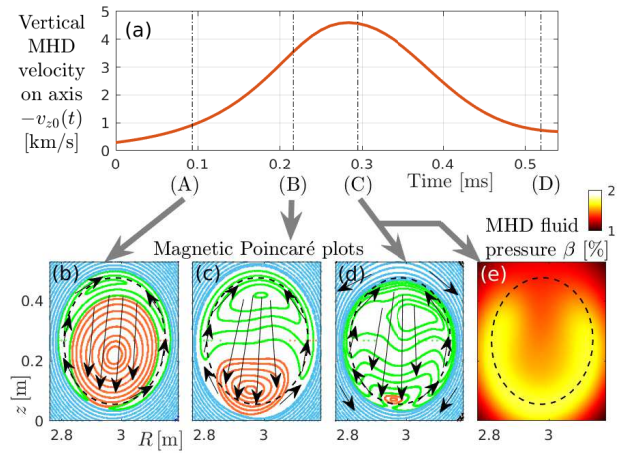} 
\caption{Simulated sawtooth crash starting from the equilibrium in Fig.~\protect\ref{fig:01_equil}. (a) Evolution of the vertical component of the MHD velocity $v_{z0} = v_z(R_0,z_0,\zeta_0)$ evaluated at the plasma center (magnetic axis) at $R_0 = 3\,{\rm m}$, $z_0 = 0.26\,{\rm m}$, $\zeta_0 = 0$. (b--d) Poincar\'{e} plots taken in the poloidal $(R,z)$ plane at $\zeta_0$, showing the topology of the magnetic field for the snapshots labeled (A), (B) and (C), whose times $0.1$, $0.2$ and $0.3\,{\rm ms}$ are measured from the instant where macroscopic displacement becomes visible. Red Poincar\'{e} contours have helicities $q < 1$. Green islands and blue periphery have $q > 1$. Arrows roughly indicate $\ExB$ flow directions. See Figs.~\ref{fig:s03_evol_enr_alpha} and \ref{fig:s06_poin} of the supplementary material for further details. (e) Contour plot of the bulk plasma beta $\beta(R,z,\zeta_0)$ at $0.3\,{\rm ms}$, snapshot (C). Here, $\beta$ behaves as an MHD fluid. Its initial form was shown in Fig.~\protect\ref{fig:01_equil}(b). The dashed circle is the initial $q=1$ surface. \vspace{-0.15cm}}
\label{fig:03_mega_saw}%
\end{figure}

Figure~\ref{fig:03_mega_saw}(a) shows the evolution of the vertical ($z$) component of the local MHD velocity $v_{z0}(t)$ measured at the magnetic axis, which consists primarily of electric drift, ${\bm v}_E = {\bm E}\times{\bm B}/B^2$. Its magnitude reaches nearly $5\,{\rm km/s}$, or $0.06\%$ of the Alfv\'{e}n speed. Here the sign of $v_{z0}$ is negative, but we have inverted it in panel (a) to visualize the growth, saturation and decay of the internal kink mode.

Magnetic reconnection converts magnetic to kinetic energy. The kinetic energy carried by the flow in Fig.~\ref{fig:03_mega_saw}(a) increases exponentially at first and saturates when the reconnection stops. Here, the reconnection phase lasts about $0.3\,{\rm ms}$ as one can see in panels (b)-(d) of Fig.~\ref{fig:03_mega_saw}, where we show Poincar\'{e} plots of the magnetic field topology during that period. The structures in Fig.~\ref{fig:03_mega_saw} should be imagined as winding helically around the plasma center, like the field line sketched in Fig.~\ref{fig:01_equil}(a). In the $\zeta_0 = 0$ plane that we use for all our Poincar\'{e} and contour plots in Fig.~\ref{fig:03_mega_saw}, the electric drift (black arrows) advects the central region of the plasma downward and returns upwards along the sides. Magnetic flux surfaces are torn and reconnected  at the bottom of the plot, where the original plasma core with $q < 1$ (red contours) shrinks and nearly vanishes around the time of snapshot (C) at $0.3\,{\rm ms}$. The field helicity in the region with green contours is slightly above unity ($q \gtrsim 1$) and nearly uniform, so the green structures appearing in the Poincar\'{e} plots may be ignored.

The dynamics seen here resemble a reconnecting internal kink mode as envisioned in the Kadomtsev model \cite{Kadomtsev75}, which has been observed in well-diagnosed experiments \cite{Park19}, although other behavior is possible (e.g., \cite{Nishimura99, Beidler11, Yu14, Jardin15, Smiet20}). The pair of convective cells associated with the internal kink redistributes the bulk plasma (modeled as an MHD fluid) in such a way that the pressure peak of Fig.~\ref{fig:01_equil}(c) acquires a horse-shoe-like structure at the end of the crash as shown in Fig.~\ref{fig:03_mega_saw}(e) \cite{Nicolas13}. (We suspect that the associated changes in the MHD force balance are responsible for the compression of the reconnected core in Fig.~\ref{fig:03_mega_saw}(d).)

While the $\ExB$ flows decay after snapshot (C), they cause further interchange and mixing as described by Wesson \cite{Wesson86} for about $0.1\,{\rm ms}$. As the flows weaken towards the end of the simulation ($\gtrsim 0.5\,{\rm ms}$), their effect is eventually overcome by sources, which tend to restore the original plasma current profile in our simulation model.

\begin{figure}
[tbp]
\centering\vspace{-0.25cm}
\includegraphics[width=0.45\textwidth]{\figures/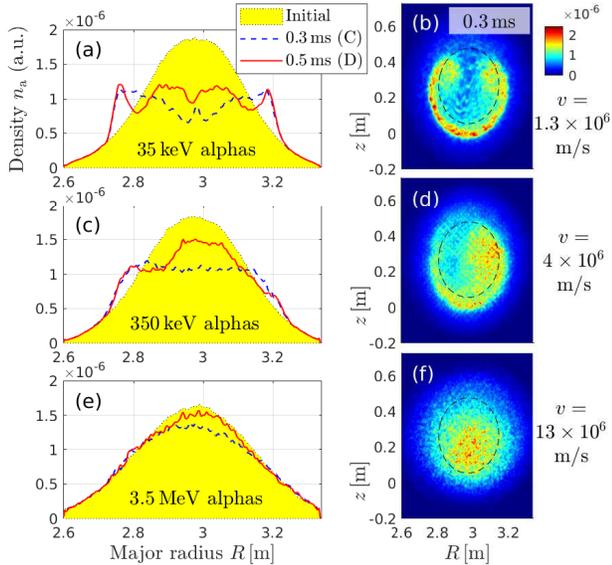}
\caption{Spatial transport of alpha particles with kinetic energies $K = 35\,{\rm keV}$ (a,b), $350\,{\rm keV}$ (c,d), $3.5\,{\rm MeV}$ (e,f) during the sawtooth crash in Fig.~\protect\ref{fig:03_mega_saw}. The left column shows snapshots of the radial density profile $n_{\rm a}(R)$ at the height of the midplane ($z_0 = 0.26\,{\rm m}$), integrated over $\zeta$ and all pitch angles. On the right, we show the density field $n_{\rm a}(R,z,\zeta_0)$ in the poloidal plane around the toroidal angle $\zeta_0 = 0$ at the time of snapshot (C) $0.3\,{\rm ms}$. The particle velocities $v = \sqrt{2K/M_{\rm a}}$ are shown for reference. These data were obtained in simulations of the reduced domain encircled by a red line in Fig.~\protect\ref{fig:01_equil}(b). Very similar results are obtained for the full domain as shown in Fig.~\protect\ref{fig:s03_evol_enr_alpha}(c) of the supplementary material.\vspace{-0.15cm}}
\label{fig:04_mega_kin_alpha}%
\end{figure}

\begin{figure}
[tbp]
\centering\vspace{-0.25cm}
\includegraphics[width=0.45\textwidth]{\figures/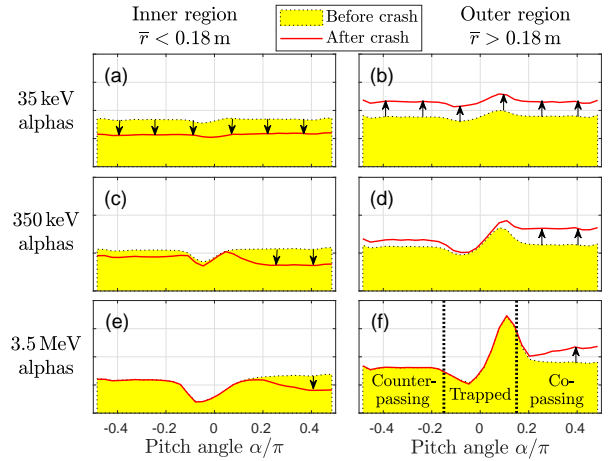}
\caption{Pre- and post-crash distributions in pitch angle $\alpha = \sin^{-1}(v_\parallel/v)$ for alpha particles with initial energies $K = 35\,{\rm keV}$ (a,b), $350\,{\rm keV}$ (c,d) and $3.5\,{\rm MeV}$ (e,f). These data were obtained in simulations of the entire plasma in order to capture all orbits, including those near the trapped-passing boundary, which traverse both the plasma core and the periphery \protect\cite{Bierwage21b}. The spatial integration is performed separately for the inner region $\overline{r} < 0.18\,{\rm m}$ (left) and the outer region $\overline{r} > 0.18\,{\rm m}$ (right), which are separated roughly by the initial $q = 1$ radius $\overline{r}_1 \approx 0.19\,{\rm m}$. More complete views of the velocity distributions are shown in Figs.~\protect\ref{fig:s08_pitch} and \protect\ref{fig:s09_enr-pitch} of the supplementary material.\vspace{-0.15cm}}
\label{fig:05_pitch}%
\end{figure}

\subsection{Alpha particle transport}

The sawtooth crash in our simulation lasts less than $0.5\,{\rm ms}$. This lies in the range of experimentally observed crash times $\tau_{\rm crash} \sim 0.1...2\,{\rm ms}$ in Fig.~\ref{fig:02_jet_saw_raw}, so we take this to be a meaningful scenario to study the redistribution of fast alphas and helium ash during such a relaxation event in JET geometry.

We fill our simulation domain with mono-energetic alphas populating all pitch angles, except for the loss cone \cite{Bierwage21b}. The initial radial profiles of the particle density at the height of the midplane ($z_0 = 0.27\,{\rm m}$) are shown as yellow shaded areas in Fig.~\ref{fig:04_mega_kin_alpha}. The widths of the profiles are chosen to be similar to the neutron emission profile observed in JET \cite{Nocente20, Kazakov21,Stancar21}, but vary somewhat due to the energy-dependence of magnetic drifts and gyroradii. Our simulations are run with a negligibly low alpha particle density to ensure that our alphas remain passive, so that the bulk plasma dynamics are identical in all cases.

Panels (a), (c) and (e) of Fig.~\ref{fig:04_mega_kin_alpha} show the evolution of the density profile of alpha particles with kinetic energies $K = M_{\rm a} v^2/2 = 35\,{\rm keV}$, $350\,{\rm keV}$ and $3.5\,{\rm MeV}$ during the simulated sawtooth crash. Around the time of snapshot (C) at $0.3\,{\rm ms}$, where the MHD flows are strongest, the central density of $35\,{\rm keV}$ alphas is temporarily reduced to $40\%$ of its initial value. It recovers partially during the aftermath of the sawtooth crash and settles at about 60\% of its initial value around the time of snapshot (D) at $0.5\,{\rm ms}$, which can be regarded as the relaxed state. In contrast, the $3.5\,{\rm MeV}$ alpha density remains centrally peaked, and recovers almost fully after a small temporary reduction. Intermediate behavior is seen at intermediate energies (see also Fig.~\ref{fig:s09_enr-pitch} of the supplementary material).

\begin{figure*}
[tbp]
\centering\vspace{-0.5cm}
\includegraphics[width=0.9\textwidth]{\figures/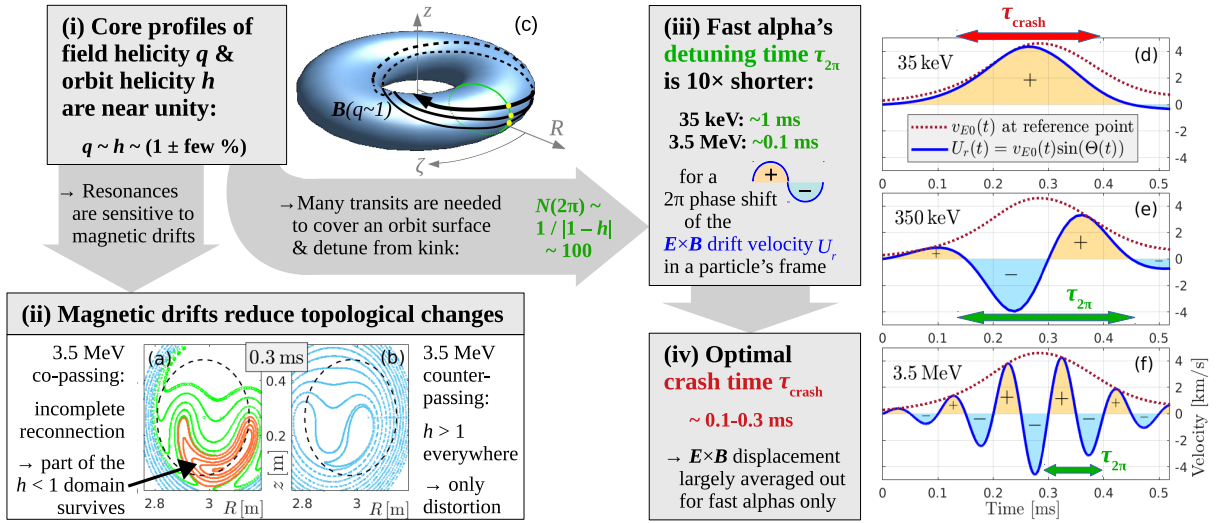}
\caption{The selective confinement of fast alpha particles observed in our simulations is due to a synergy between the effects of (i) near-unity helicities and (ii) magnetic drifts, and the time scales for (iii) resonance detuning and (iv) the sawtooth crash.\vspace{-0.15cm}}
\label{fig:06_synergy}%
\end{figure*}

The redistribution of particles at different pitch angles $\alpha = \sin^{-1}(v_\parallel/v)$ with $v_\parallel/v  \equiv {\bm v}\cdot{\bm B}/(vB)$ is shown in Fig.~\ref{fig:05_pitch}. Co-passing particles occupy roughly the domain $\alpha \gtrsim 0.15\pi$, counter-passing particles have $\alpha \lesssim -0.15\pi$, and particles trapped by the magnetic mirror force are found around $-0.15\pi \lesssim \alpha \lesssim 0.15\pi$. Here, the attributes `co' and `counter' refer to the direction of the plasma current, which in the present case coincides with the direction of ${\bm B}$. The initial particle distributions (yellow-shaded areas in Fig.~\ref{fig:05_pitch}) have been prepared to be as uniform in pitch as is physically possible. The nonuniformity around $\alpha \approx 0$ that increases with energy $K$ is an inevitable consequence of the sharply peaked density profile, magnetic drifts and the presence of many non-standard orbits \cite{Bierwage21b}.

Results for the slow $35\,{\rm keV}$ alphas are plotted in Fig.~\ref{fig:05_pitch}(a,b), where one can see that the sawtooth crash causes net outward displacement of nearly uniform magnitude across the entire range of pitch angles $-\pi/2 \leq \alpha \leq \pi/2$. At $350\,{\rm keV}$ in Fig.~\ref{fig:05_pitch}(c,d), about 40\% of the co-passing alphas are displaced, whereas counter-passing particles are only weakly perturbed, and mirror-trapped particles even less. In the case of $3.5\,{\rm MeV}$ in Fig.~\ref{fig:05_pitch}(e,f), net transport is observed only in the domain of co-passing particles, whose number in the inner domain decreases by about $20\%$. The small reduction in the overall density in Fig.~\ref{fig:04_mega_kin_alpha}(e) is due to these displaced co-passing particles.

\subsection{Underlying synergism}

An energy dependence of sawtooth crash-induced transport as seen in Fig.~\ref{fig:04_mega_kin_alpha} has been predicted by Kolesnichenko \& Yakovenko \cite{Kolesnichenko96}, as has been the fact that the threshold energy is lower for trapped particles than for passing ones as seen in Fig.~\ref{fig:05_pitch}. The reason for the observed difference between co- and counter-passing particles in Fig.~\ref{fig:05_pitch}, however, remained elusive \cite{Jaulmes14, Geiger15}. Here we propose an explanation and a complete physical picture.

A schematic illustration of the four physical factors whose synergistic interplay is responsible for the energy-selective confinement of fast alphas in our simulations is shown in Fig.~\ref{fig:06_synergy}. Before going into the details, here is a quick run-through: The first key factor, represented by box (i) in Fig.~\ref{fig:06_synergy}, is that the field helicity $q$ is close to unity. This has the consequence that (ii) alpha particle resonances with respect to the internal kink are sensitive to magnetic drifts, and (iii) only fast alphas have a resonance detuning time $\tau_{2\pi} \sim 0.1\,{\rm ms}$ that is shorter than (iv) the typical sawtooth crash time $\tau_{\rm crash} \sim 0.3\,{\rm ms}$. Hence:
\begin{itemize}
\item  When $q \sim 1$ and magnetic drifts are large, there exists an optimal combination of time scales, where resonant interactions between the internal kink and fast alphas can be largely avoided; thus, facilitating good confinement of fast alphas only.
\end{itemize}

\noindent Factors (i), (iii) and (iv) were anticipated by theoretical analyses in \cite{Kolesnichenko96}. The physical picture is completed here by including (ii) the effect of magnetic drifts on the resonances. Now, let us elucidate this synergism in detail.

The condition for a charged particle to resonate with the internal kink can be expressed as $h \equiv \omega_{\rm tor}/\omega_{\rm pol} = 1$, where the orbit helicity $h \approx \overline{r}v_{\rm tor}/(R_0 v_{\rm pol})$ is the ratio of toroidal and poloidal transit frequencies and can be thought of as the kinetic counterpart of the field helicity $q \approx \overline{r}B_{\rm tor}/(R_0 B_{\rm pol})$ modified by the combined effect of magnetic drifts ${\bm v}_{\rm d}$ and the mirror force \cite{Shinohara18, Shinohara20}.

The radial excursion caused by magnetic drifts is given by $(\Delta R)_{\rm d} \approx v_{\rm d}/\omega_{\rm pol} \approx v_\parallel[1 + v_\perp^2/(2v_\parallel^2)] M/(Qe B_{\rm pol} \overline{r})$, so it is proportional to a particle's mass-to-charge ratio $M/Q$ and velocity $v$, and inversely proportional to the plasma current \cite{MiyamotoBook3}. Co-(counter-)passing orbits are shifted out-(in-)ward in major radius $R$.

\begin{figure}
[tbp]
\centering\vspace{-0.25cm}
\includegraphics[width=0.45\textwidth]{\figures/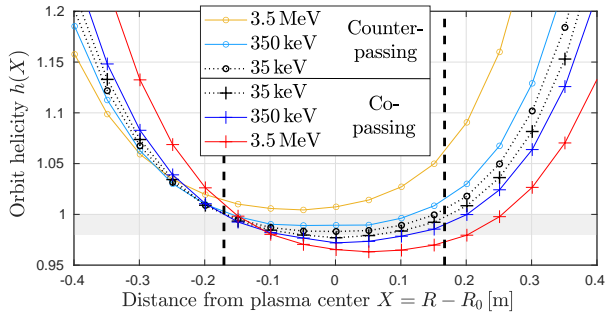}
\caption{Profiles of the orbit helicity $h(X) = \omega_{\rm tor}/\omega_{\rm pol}$ plotted as a function of the radial position $X = R - R_0$ where an alpha particle orbit crosses the midplane $z_0 = 0.26$. Dashed vertical lines indicate the $q=1$ radii for the $q$ profile in Fig.~\protect\ref{fig:01_equil}. Orbits with initial helicities in the range ${\rm min}\{q\} \lesssim h \lesssim 1$ (shaded area) and nearby are likely to be subject to resonant interaction and `reconnection' some time during the sawtooth crash (an accurate prediction may be made using Hamiltonian analysis in helical coordinates as, for instance, in Fig.~8.12 of \protect\cite{WhiteTokBook3}). The pitch dependence of the outer $h=1$ radius for $3.5\,{\rm MeV}$ alphas is shown in Fig.~\protect\ref{fig:s07_jet_3500keV_resonance_X-r-pitch} of the supplementary material.}\vspace{-0.15cm}
\label{fig:07_helicity}%
\end{figure}

Figure~\ref{fig:07_helicity} shows a few examples of orbit helicity profiles $h(X)$ with $X = R - R_0$ for co- and counter-passing alpha particles with energies $K = 35\,{\rm keV}$, $350\,{\rm keV}$ and $3.5\,{\rm MeV}$, and velocity pitch $v_\parallel/v = \sin(\pm 0.48\pi) \approx \pm 1$. One can see that, with increasing kinetic energy $K$, the orbit helicity profiles $h$ deviate increasingly from the field helicity profile $q \approx h(35\,{\rm keV})$. The mirror force causes particles to spend more time in regions of smaller $R$, where the field is stronger, so that the orbit helicity $h$ is reduced for the outward-shifted co-passing ions, and increased for inward-shifted counter-passing ions. Figure~\ref{fig:07_helicity} shows that in the present case these effects are large enough to entirely eliminate the $h=1$ resonance for counter-passing $3.5\,{\rm MeV}$ alphas by raising their $h$ profile above unity everywhere.

These differences between the field and orbit helicities have two consequences. First, the parallel electric field $E_\parallel = {\bm E}\cdot{\bm B}/B$ in the magnetic reconnection layer near $q = 1$ has less influence on faster ions. Second, magnetic drifts reduce the amount of reconnection that occurs in the topology of fast alpha particle orbits as highlighted in box (ii) of Fig.~\ref{fig:06_synergy}. Panel (a) shows that reconnection in co-passing orbit topology is incomplete, so that a large region with $h < 1$ survives through the crash. For counter-passing orbits in panel (b), we have always $h > 1$ everywhere, which means that an $h=1$ resonance never forms and the orbit topology is merely distorted by the magnetic perturbations associated with the internal kink. More details about the orbit topology evolution can be found in Fig.~S6 of the supplementary material.

This mechanism explains the unresponsiveness of counter-passing alphas and partial mixing of co-passing alphas in Fig.~\ref{fig:05_pitch}. However, it must be noted that the observed amount of co-/counter-passing asymmetry is linked to the choice of the boundary between the `inner' and `outer' regions, which we have placed near the initial $q=1$ radius $\overline{r}_1 \approx 0.19\,{\rm m}$. For instance, counter-passing particles in the intermediate case with $350\,{\rm keV}$ are also subject to mixing, but this is only partially visible in Fig.~\ref{fig:05_pitch}(c,d) because their mixing radius is smaller than the $q = 1$ radius, as is evident from Fig.~\ref{fig:07_helicity}. Results for a smaller `inner' region ($\overline{r} < 0.06\,{\rm m}$) are shown in Fig.~\ref{fig:s09_enr-pitch}(c-e) of the supplementary material.

Let us now proceed to the next physical factor. The second implication of having $q \sim h \sim 1$ in box (i) of Fig.~\ref{fig:06_synergy} is that field and orbit trajectories nearly close on themselves after one toroidal transit as illustrated schematically in panel (c). Taking a concrete value for illustration, say $h \sim 1 \pm 0.01$, this means that a particle has to perform on the order of $N(2\pi) \sim 1/|1-h| \sim 100$ toroidal transits before its trajectory covers a toroidal orbit surface. When viewed from above, the motion illustrated schematically in Fig.~\protect\ref{fig:06_synergy}(c) resembles the slow apsidal precession of a planetary orbit around the Sun.

During these $N(2\pi)$ transits, the phase of the radial ${\bm E}\times{\bm B}$ drift velocity $U_r(t)$ measured in the frame of reference moving with a chosen particle will also complete a $2\pi$ phase shift. This is illustrated in panels (d)--(f) of Fig.~\ref{fig:06_synergy}. In our simplified example, we assume the that the electric drift in the moving frame has a sinusoidal form $U_r(t) = v_{E0}(t)\sin(\Theta(t))$. For the envelope $v_{E0}(t)$ we chose the vertical MHD velocity from Fig.~\ref{fig:03_mega_saw}(a) as $v_{E0} = -v_{z0}$. The evolution of the phase is modeled as $\Theta(t) = 2\pi t/\tau_{2\pi}$, where $\tau_{2\pi} = \tau_{\rm tor}N(2\pi) \sim \tau_{\rm tor}/|1 - h|$ is the time scale for the kink mode's phase to slip by $2\pi$. Thus, $\tau_{2\pi}$ plays the role of a resonance detuning time.

As indicated in box (iii) of Fig.~\ref{fig:06_synergy}, the resonance detuning times of newly born $3.5\,{\rm MeV}$ alphas and $35\,{\rm keV}$ helium ash differ by a factor 10. Incidentally, the typical sawtooth crash time $\tau_{\rm crash}$ of a few $100\,\mu{\rm s}$ happens to be shorter than the $1\,{\rm ms}$ detuning time of $35\,{\rm keV}$ alphas and longer than the $0.1\,{\rm ms}$ detuning time of $3.5\,{\rm MeV}$ alphas in our setup. This leads to the final box (iv) of Fig.~\ref{fig:06_synergy}: only for sufficiently fast alphas the net ${\bm E}\times{\bm B}$ displacement is likely to vanish through the cancellation of positive and negative peaks of $U_r(t)$ as in panel (f). Depending on the initial phase, there is still a 50\% chance for cancellation at $350\,{\rm keV}$ in panel (e), where $\tau_{2\pi} \sim \tau_{\rm crash}$. In contrast, the $35\,{\rm keV}$ alphas in panel (d) typically remain in phase with the kink's electric field for the entire duration of the crash. This explains why the density field of slow $35\,{\rm keV}$ alphas in Fig.~\ref{fig:04_mega_kin_alpha}(b) develops the same horse-shoe-like structure as the MHD fluid in Fig.~\ref{fig:03_mega_saw}(e). At the intermediate energy of $350\,{\rm keV}$ in Fig.~\ref{fig:04_mega_kin_alpha}(d), the density field has a swirling tear-drop structure as the poloidal spreading competes with the displacement due to the $\ExB$ drift on a similar time scale. At $3.5\,{\rm MeV}$, the toroidal speed is so high that poloidal spreading outpaces $\ExB$ drifts, so the density field in Fig.~\ref{fig:04_mega_kin_alpha}(f) maintains a compact peak with only a minor helical distortion.

We emphasize once more that the magnetic drift effect in box (ii) of Fig.~\ref{fig:06_synergy} is crucial. For instance, the rapid motion of thermal electrons allows them to satisfy conditions (iii) and (iv) in Fig.~\ref{fig:06_synergy}, but the lack of magnetic drifts causes them to stick closely to magnetic field lines and undergo strong mixing during sawtooth crashes that involve magnetic reconnection. This has been verified by varying the mass-to-charge ratio of our simulation particles as described in the supplementary material (Fig.~\ref{fig:s05_mega_kin_10mass10}).

Finally, we note that Fig.~\ref{fig:06_synergy} shows a simplified representation of reality. It is only meant to convey the basic concepts. For instance, the transit number $N(2\pi)$ that determines the detuning time $\tau_{2\pi}$ varies with particle energy and radial location as one can readily infer from the orbit helicity profiles $h(X)$ in Fig.~\ref{fig:07_helicity}. These $h$ profiles also evolve in time, with a tendency to rise during the crash. Thus, the resonances are dynamic: their location, width and their very existence evolve rapidly. Moreover, the $\ExB$ flow pattern is nonuniform in space. Along particle orbits that perform large magnetic drifts, the direction of the electric drift may thus vary even during a single transit. This reduces the effective magnitude of the electric drifts for fast alphas. However, this well-known orbit-averaging effect alone is not sufficient to prevent strong mixing as can be verified from the particle mass scan reported in the supplementary material (Fig.~\ref{fig:s05_mega_kin_10mass10}). The selective confinement of fast alphas in a wide range of pitch angles is realized only through the synergistic effect that arises from the four factors in Fig.~\ref{fig:06_synergy} combined.

\section{Discussion}

Simulations of a large tokamak plasma based on JET confirmed the existence of a parameter window where the majority of MeV-class alpha particles can remain well-confined in the plasma core during a benign sawtooth crash that strongly redistributes less energetic ions and electrons. We proposed a physical picture that extends the existing theory \cite{Kolesnichenko96} by accounting for the modification of alpha particle resonances \cite{Kolesnichenko98, Kolesnichenko00a, Kolesnichenko00b, Teplukhina21} and orbit topology via magnetic drifts. This explains the observed differences between the responses of co- and counter-passing alphas during a sawtooth crash \cite{Jaulmes14, Geiger15, Teplukhina21}. A reduced model that captures these effects is now available \cite{Podesta22}.

The insights won are of interest for the tokamak-based fusion reactor R\&D programs that are currently pursued around the world. Those R\&D activities would benefit from the possibility of using sawteeth for removing helium ash and other impurities without deteriorating fast alpha confinement. All else being similar, the time scales summarized in Fig.~\ref{fig:06_synergy} would be 2--3 times longer in ITER and DEMO reactors due to their larger major radii $R_0 \approx 6\,{\rm m}$ and $9\,{\rm m}$, respectively. The scenario we studied hence lies quantitatively in the right `ball park'. Thus motivated, we conclude this study with a preliminary discussion about the practicality of the method.

Overall, it has become clear that the parameter window of interest is narrow. This poses several practical obstacles. First, since the magnetic drifts play an important role, the mechanism described here may be utilizable only when the plasma current is not too high. The advantages and disadvantages of this operational regime will have to be weighed. Potential applications in spherical tokamaks may be worthy of consideration \cite{Bakharev15, Liu18, Jackson20}.

Second, the underlying theoretical picture implies that the redistribution of the alpha particles is sensitive to the pre-crash profile of the field helicity $q$. This prediction is corroborated by preliminary results of $q$ profile scans reported in the supplementary material, which indicate that the fast alpha profiles undergo significant flattening when $|1 - q| \gtrsim \O(5\%)$ for the considered plasma parameters ($I_{\rm p} = 2.5\,{\rm MA}$, $B_0 = 3.5\,{\rm T}$). In order to utilize the hoped-for energy-selective mixing and confinement in a reactor, it is thus necessary to ensure that only certain types of sawteeth occur, namely those for which the field helicity remains close to unity ($q \sim 1$). Provided that this is practically possible, it requires both a better understanding of sawtooth physics as pursued in \cite{KumarAPS21} and precise plasma control schemes, where one controls not only the sawtooth period but also the form of the crash.

In this context, it is also necessary to quantify the stabilizing or destabilizing effect that fast alphas in realistic concentrations exert on the internal kink and other MHD modes \cite{Porcelli91, Porcelli92, Porcelli94, Kiptily21}. The same counts for the influence of sawtooth-induced alpha particle transport on the background plasma from the viewpoints of heating, current drive and plasma rotation. Investigations in this direction are motivated by the local imbalance between co- and counter-passing alphas that was caused by the sawtooth crashes in our simulations. This effect can be significant even for larger crashes triggered at $|1 - q| \sim \O(10\%)$ and may be a subject of interest on its own. It can influence the evolution of the $q$ profile by adding negative and positive plasma current, and it may exert a sheared toroidal torque around the radius of the pre-crash $q = 1$ surface. This, in turn, can lead to complex nonlinear feedback whose consequences deserve further study.

Last but not least, another potential obstacle becomes evident if one considers not only a single sawtooth crash but multiple sawtooth cycles. Since a sawtooth crash is a mixing process and, thus, does not affect an already flat profile, it is clear that the mechanism we have described becomes noticeable only if the density profiles of the fusion-born alphas and resulting ash are peaked within the mixing radius. The formation of such a peaked profile in the density of fusion-born alphas requires that the D-T fusion fuel has a peaked pressure profile in the first place. The sustained core-localized alpha heating can help to recover this condition after a sawtooth crash, but the interval between successive sawteeth must be sufficiently long to allow for the recovery of a sharply peaked plasma pressure. Although it is still unclear how to realize this situation in practice, the observation of impurity accumulation in tokamak plasmas with enhanced overall confinement suggests that there exists a `density pinch' mechanisms that can lead to central peaking and requires further study. Techniques to deliver fusion fuel deep into the plasma core would also help and constitute yet another challenging topic of fusion research.

On a positive note, the practical considerations and the physics we have discussed exemplify the rich nonlinear dynamics that magnetically confined fusion plasmas support. We expect that successful ITER experiments have not only the potential to yield valuable insights needed for fusion power plants; ITER is also likely to inspire new experiments for burning plasma physics studies, with sawtooth crashes and the associated reconnection in magnetic and orbit topology playing a prominent role. The parameter window of interest that we identified in this work --- namely, a field helicity profile $q(\overline{r})$ close to unity --- lies precisely in the regime envisioned for ITER long-pulse operation. At the same time, this regime still exhibits unsolved mysteries and, thus, opportunities (e.g., \cite{Guenter01, Joffrin02, Misguich02, Chu06, Jardin15}). The effects discussed here and further research should throw more light on these matters.

\begin{table}[b]
\begin{tabular}{lrcl}
\hline
Major radius & $R_0$ & = & $3.0\,{\rm m}$ \\
Mean minor radius & $\left<a\right>$ & $\approx$ & $1.2\,{\rm m}$ \\
Plasma current & $I_{\rm p}$ & $=$ & $2.5\,{\rm MA}$ \\
Toroidal field strength & $B_0$ & $=$ & $3.7\,{\rm T}$ \\
Thermal/magnetic pressure ratio & $\beta_0$ & $=$ & $2\%$ \\
Safety factor (magnetic axis) & $q_0$ & $=$ & $0.98$ \\
Safety factor (last closed flux surface) & $q_a$ & $=$ & $5.44$ \\
Number density (bulk, deuterium) & $n_{{\rm b}0}$ & $=$ & $5.3\times 10^{19}\,{\rm m}^{-3}$ \\
Number density (alpha particles) & $n_{{\rm a}0}$ & $\lesssim$ & $10^{-7}\times n_{{\rm b}0}$ \\
\hline
\end{tabular}
\caption{Plasma parameters in simulations based on JET pulse number 95679. The subscript 0 indicates that a quantity is measured at the center (magnetic axis) of the plasma. The alpha particle density in the simulations was initialized with a negligibly low value, so that they behaved as passive tracer particles that did not affect the evolution of MHD modes.}
\label{tab:parm}
\end{table}

\section*{Acknowledgments}

A.B.\ thanks Yasushi Todo (NIFS, Japan) for valuable support in connection with the code {\tt MEGA}. Insightful discussions with Nikolai Gorelenkov, William Heidbrink and Yurii Yakovenko following the presentation of our results at the 17th IAEA TCM EPPI meeting (Dec.\ 2021) are thankfully acknowledged. The simulations reported here were carried out using the supercomputer JFRS-1 at Computational Simulation Centre of International Fusion Energy Research Centre (IFERC-CSC) in Rokkasho Fusion Institute of QST. Preliminary studies were also conducted using the supercomputer SGI ICE X in the Japan Atomic Energy Agency (JAEA).

The work by S.S.\ was supported by Grants-in-Aid for Scientific Research from JSPS (Grant No.\ 20K14447). This work has been partially carried out within the framework of the EUROfusion Consortium and has received funding from the Euratom research and training programme 2014-2018 and 2019-2020 under Grant Agreement No.\ 633053. The views and opinions expressed herein do not necessarily reflect those of the European Commission.

\section{Methods}

\subsection{Code availability}

Further information concerning the hybrid code {\tt MEGA} \cite{Todo03b,Todo05,Todo98} and the version used in this work \cite{Bierwage16c, Bierwage17a} can be made available by the corresponding author upon reasonable request. The source code may be obtained after establishing an official research collaboration agreement with National Institutes for Quantum Science and Technology (QST).

\subsection{Simulation model and parameters}

A detailed description of the model equations and numerical techniques along with benchmarks, convergence tests, sensitivity studies and experimental validation for different types of fast-ion-driven instabilities in tokamaks can be found in the literature \cite{Bierwage17a, Bierwage17c, Bierwage19, Bierwage16c, Todo15, Bierwage18}. The simulation setup used in the present work was described in \cite{Bierwage21b} and further details including sensitivity tests are provided in the supplementary material accompanying this paper. For the reader's convenience, the following paragraphs provide a summary of the numerical parameters used in the simulations and data analysis. A summary of relevant plasma parameters is given in Table~\ref{tab:parm}.

\subsection{Discretization, dissipation, filtering}

Our cylindrical mesh in the reduced simulation domain bounded by the red circle in Fig.~\ref{fig:01_equil}(b) consists of $N_R\times N_z\times N_\zeta = 220\times 220\times 96$ points; very similar results were obtained with $520\times 520\times 96$ and $220\times 220\times 192$. Simulations for the full domain used $520\times 520\times 96$ grid points. The 4th-order Runge-Kutta time step was about $1\,{\rm ns}$ for the MHD solver, which is constrained by the Courant-Friedrichs-Lewy (CFL) condition for fast magnetosonic waves. The value of the specific heat ratio that controls the plasma compressibility in our MHD simulations was fixed at $\Gamma = 5/3$. Larger time steps can be used for the alpha particles: at $35\,{\rm keV}$ it suffices to advance particles with a 40 times greater time step, whereas $3.5\,{\rm M}$ alphas were pushed every 4th MHD time step. Test runs with more frequent pushing produced the same results.

The number of simulation particles representing the guiding centers of mono-energetic populations of alpha particles lay in the range $(1.5...4)\times 10^6$ in the reduced domain and $(8...22)\times 10^6$ in the full domain. A full-domain simulation with alphas distributed uniformly over the energy range $(0.35...3.5)\,{\rm MeV}$ used $37\times 10^6$ simulation particles (see Fig.~\ref{fig:s09_enr-pitch} of the supplementary material).

Our simulations were run with 4-point gyroaveraging around a particle's guiding center \cite{Bierwage16c}, although very similar results can be obtained in the zero-Larmor-radius limit, even for $3.5\,{\rm MeV}$ alpha particles. We expect that the effect of gyroaveraging will be more important when the population of alpha particles is large enough to influence MHD modes, which was not the case here.

The values of the MHD coefficients controlling resistive, viscous and thermal diffusion are $\eta/\mu_0 = \nu = \chi = 10^{-6} v_{\rm A0} R_0$, which is considered to be a reasonable compromise between numerical and physical considerations \cite{Bierwage16c}. In the full domain, we filtered out toroidal harmonics $\exp(in\zeta)$ with $|n| > 12$ to suppress numerical instabilities. Such filtering was also necessary in $q$ profile scans where we simulated 50\% of the magnetic flux space (Figs.~\ref{fig:s10_evol_enr-q_scan24}-\ref{fig:s13_scan24_pitch_3500keV} of the supplementary material), but it was not necessary when simulating only the inner 25\% of the magnetic flux space for our default case with $q_0 = 0.98$.

\subsection{Poincar\'{e} analysis of field and orbit topology}

All Poincar\'{e} plots presented in this work --- including Figs.~\ref{fig:03_mega_saw} and \ref{fig:06_synergy} of the main article and Figs.~\ref{fig:s04_poin_mhd-full_islands}, \ref{fig:s06_poin}, \ref{fig:s11_scan2_poin} and \ref{fig:s12_scan4_poin} of the supplementary material --- were obtained by tracing magnetic field lines or the guiding center orbits of alpha particles in the self-consistent perturbed magnetic field ${\bm B}$ of the hybrid simulation. Although present in the simulations, electric drifts were not included in the Poincar\'{e} analysis and the magnetic field was not evolved while the test particles traced out their orbits. This means that our Poincar\'{e} contours reflect only the {\it instantaneous topology} of the orbits, not the motion of particles in the original hybrid simulation of the sawtooth crash, where the magnetic configuration changes while a particle is tracing out a toroidal orbit surface. See also Ref.~\cite{Kolesnichenko96} for a related discussion.

The test particles were launched from the outer midplane ($R > R_0$, $z = z_0 = 0.26\,{\rm m}$) and advanced using a 4th-order Runge-Kutta algorithm. For each initial position, we recorded at least 450 Poincar\'{e} sections in the poloidal $(R,z)$ plane at $\zeta_0 = 0$. The direction in which successive points appear in the Poincar\'{e} plot allowed us to identify whether the local value of the field helicity $q$ and orbit helicity $h$ was greater or smaller than unity. This information was then used to assign colors for each of the Poincar\'{e} contours: red/orange for $q,h < 1$ and blue/green for $q,h > 1$.

\subsection{Data availability}

The data of {\tt MEGA} simulations are stored on devices administered by National Institutes for Quantum Science and Technology (QST). These data may be obtained after establishing an official research collaboration agreement with QST. The experimental data for JET pulse 95679 that is shown in Fig.~\ref{fig:02_jet_saw_raw} can be identified as (a) {\tt KK3-TE24} and (b) {\tt SXR-H10-JETPPF} (calibrated). The reference equilibrium reconstructed by {\tt TRANSP} modeling is {\tt g95679\_83-50.22554.eqdsk} (time stamp: 2019/12/19, 19:59).

\section{Supplementary material}
\setcounter{figure}{0}
\renewcommand{\thefigure}{S\arabic{figure}}
\renewcommand{\figurename}{Fig.}

\subsection*{1. Simulation scenario design}

The simulation setup is described in some detail in \cite{Bierwage21b}, where we focused on the modeling of an alpha particle distribution. Here we describe in more detail the model for the bulk plasma in our sawtooth crash simulations.

\begin{figure}
[tbp]
\centering\vspace{-0.25cm}
\includegraphics[width=0.45\textwidth]{\figures/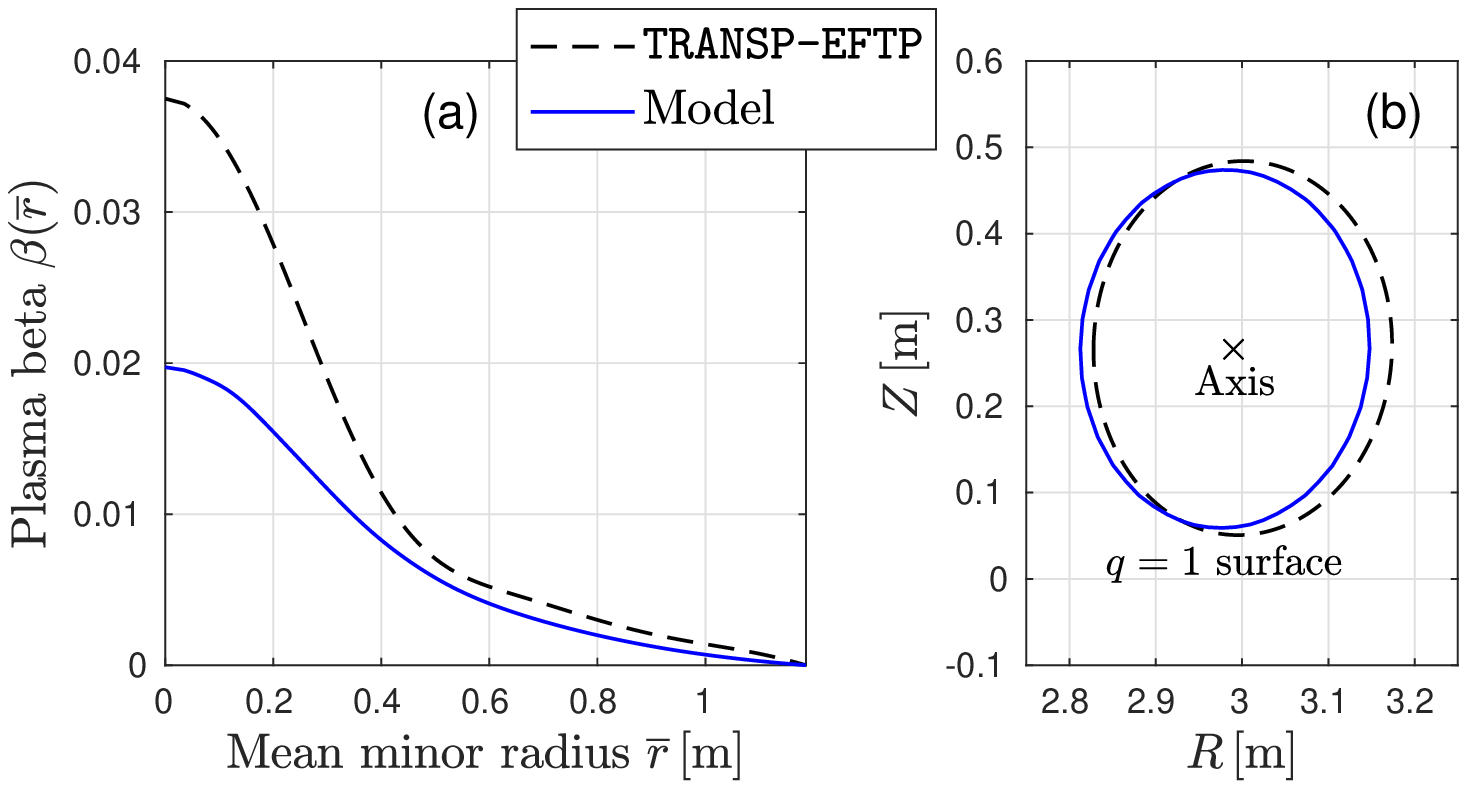}
\caption{(a) Pressure profile used for MHD force balance calculations. The dashed curve is the pressure profile predicted by integrated {\tt TRANSP-EFTP} simulation for JET pulse 95679 at 50.22554 seconds, which includes the contribution of fast ions from beams and RF heating. The solid curve is our model, which includes only the estimated bulk plasma component. (b) Excluding the fast ion component from the MHD pressure reduces the Shafranov shift, so that the $q = 1$ surface shifts inward in $R$ as shown here for the case with $q_0 = 0.98$.}\vspace{-0.15cm}
\label{fig:s01_prof_beta}%
\end{figure}

The profiles of both the fast ion pressure and the MHD pressure are based on a reconstruction of a JET plasma using the code {\tt TRANSP-EFTP} as reported in \cite{Stancar21}. The MHD pressure in these {\tt TRANSP-EFTP} simulations included the fast ion component. We have subtracted that part as shown in Fig.~\ref{fig:s01_prof_beta} and recomputed the equilibrium using only the estimated pressure of the bulk plasma.

By reducing the pressure, we were able to suppress resistive interchange modes that would otherwise overwhelm the internal kink mode in our visco-resitive MHD simulations with Lundquist and Reynolds numbers $S = {\rm Re} = 10^6$. Here, $S = \tau_\eta/\tau_{\rm A0}$ is the ratio of the resistive and toroidal Alfv\'{e}nic time scales, $\tau_\eta = \mu_0 R_0^2/\eta$ and $\tau_{\rm A0} = R_0/v_{\rm A0} \approx 0.4\,\mu{\rm s}$, where $\eta$ is the electrical resistivity. Note however that a certain amount of MHD pressure is beneficial as it seems to have a numerically stabilizing effect: test runs initialized with a zero-pressure MHD equilibrium terminated abnormally during the nonlinear saturation stage. We suspect that compressible MHD effects help to regularize the dynamics, for instance by reducing fine structures in the reconnection layer that may otherwise cause numerical problems. Possible mechanisms realizing this include secondary interchange modes and magnetosonic wave propagation. The value of the specific heat ratio that controls compressibility in our MHD simulations was fixed at $\Gamma = 5/3$.

\begin{figure}
[tbp]
\centering\vspace{-0.25cm}
\includegraphics[width=0.45\textwidth]{\figures/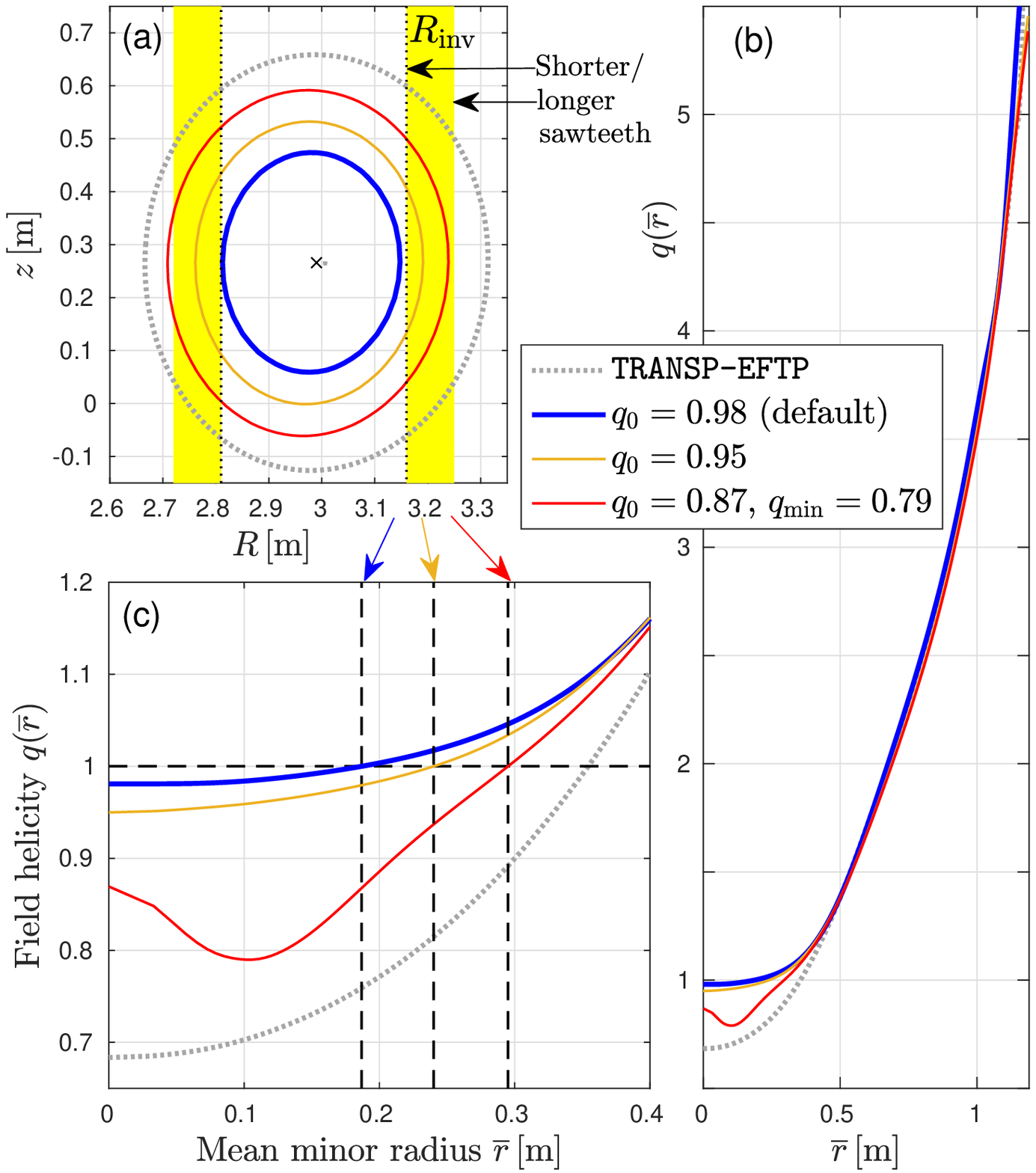}
\caption{Modeling of the field helicity (safety factor) profile $q(\overline{r})$. Panel (a) shows the contours of the $q=1$ radii in the poloidal $(R,z)$ plane for the reference equilibrium computed by {\tt TRANSP-EFTP} (dotted) and our three model profiles (solid). The yellow shaded areas indicate the approximate range of sawtooth inversion radii $R_{\rm inv}$ observed in the JET reference pulse 95679 that was used to constrain the $q=1$ radii of the model profiles (before adjusting the pressure as shown in Fig.~\protect\ref{fig:s01_prof_beta}). Panel (b) shows the full $q(\overline{r})$ profiles, whose structure in the central core is shown enlarged in panel (c).}\vspace{-0.2cm}
\label{fig:s02_prof_qsafe}%
\end{figure}

As stated in the text of the main article, we treat the $q$ profile as a free parameter. Figure~\ref{fig:s02_prof_qsafe}(a) shows that the $q=1$ radius predicted by {\tt TRANSP-EFTP} (which did not account for sawtooth activity) is much larger than the experimentally observed inversion radii $R_{\rm inv}$. Our model profiles for $q(\overline{r})$ were chosen to match the {\tt TRANSP-EFTP} profile only in the outer region of the plasma as shown in Fig.~\ref{fig:s02_prof_qsafe}(b), whereas the inner portion shown in Fig.~\ref{fig:s02_prof_qsafe}(c) was modified within the limits of $R_{\rm inv}$ inferred from experiments. The MHD equilibria for our simulations were constructed using the code {\tt CHEASE} \cite{Luetjens96}, taking as input the modeled pressure profile in Fig.~\ref{fig:s01_prof_beta}(a) and the $q$ profiles in Fig.~\ref{fig:s02_prof_qsafe}.

Besides being an essential physics ingredient for our study, the fairly flat $q$ profile in the main article's Fig.~\ref{fig:01_equil}(c) with $q \sim 1$ inside the $q = 1$ surface also has computational advantages. In combination with the suspected regularizing effect of MHD pressure mentioned in the previous paragraph, a flat $q$ profile prevents the magnetic reconnection layer from collapsing rapidly into a narrow sheet, which would be difficult to resolve numerically even with the relatively small Lundquist and Reynolds numbers $S = {\rm Re} = 10^6$ in our simulation. For instance, we were not able to perform a simulation using the original $q$ profile from the steeply rising reference equilibrium from the {\tt TRANSP-EFTP} simulation, which is shown as a dotted line in Fig.~\ref{fig:s02_prof_qsafe}(b,c) and has a low central value of $q_0 = 0.68$. A common feature of simulations that terminated abnormally is the formation of wide-spread magnetic chaos.

MHD simulations were successfully be performed with the three model $q$ profiles in Fig.~\ref{fig:s02_prof_qsafe}. The blue profile (default) with $q_0 = 0.98$ and $q=1$ radius $\overline{r} \approx 0.19\,{\rm m}$ has been studied most intensively, and the main results were reported in the main article. The other two profiles with $q_0 = 0.95$ at $\overline{r} \approx 0.24\,{\rm m}$ (orange) and $q_0 = 0.97$ at $\overline{r} \approx 0.30\,{\rm m}$ (red) are being used for sensitivity tests, preliminary results of which are presented in Section~7 below. Judging by the Alfv\'{e}n mode activity seen in spectral analyses of experimental data \cite{Kazakov21}, we speculate that our non-monotonic model $q$ profile with $q_0 = 0.87$ and $q_{\rm min} = 0.79$ (red) may be closest to the conditions that were actually present during the quasi-steady-state phase in JET pulse 95679 that we used as a reference.

As is typical for simulations like ours, the MHD model includes a source term representing the tokamak's loop voltage, which balances the resistive dissipation of the initial current profile. Without this term, our central $q_0 = 0.98$ would rise above unity on the time scale of a few $100\,\mu{\rm s}$, which implies that resistive dissipation participates in the simulated dynamics on a global scale. We emphasize again that simulating the collisionless reconnection process that is thought to occur during real Kadomtsev-type sawtooth crashes is still an active area of research  \cite{KumarAPS21}. MHD simulations like ours may thus be viewed as an attempt to mock up the micro-scale dynamics using resistive dissipation in such a way that the macro-scale dynamics are consistent with observations.

\begin{figure}
[tbp]
\centering
\includegraphics[width=0.45\textwidth]{\figures/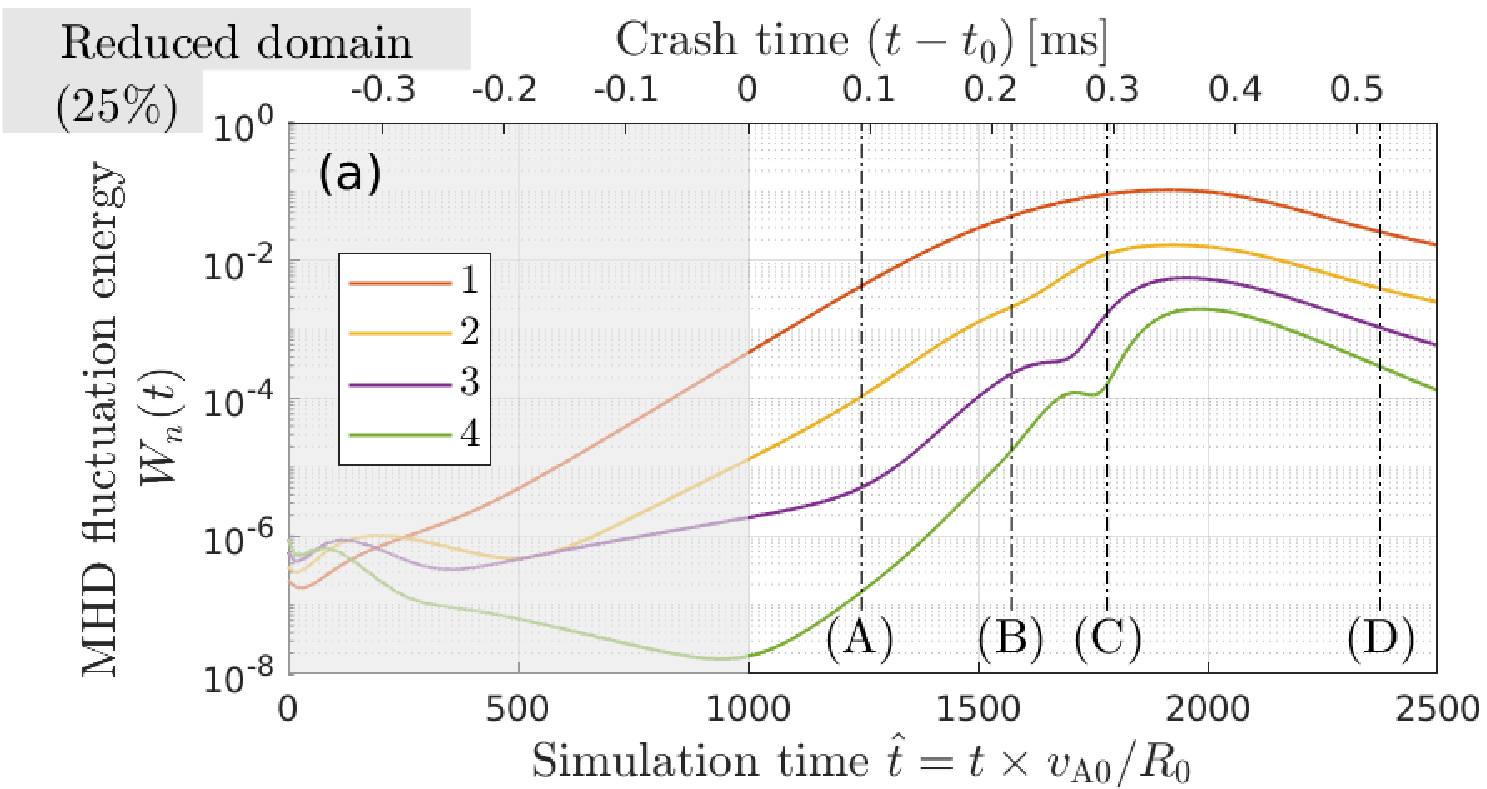} \vspace{0.2cm} \\
\includegraphics[width=0.45\textwidth]{\figures/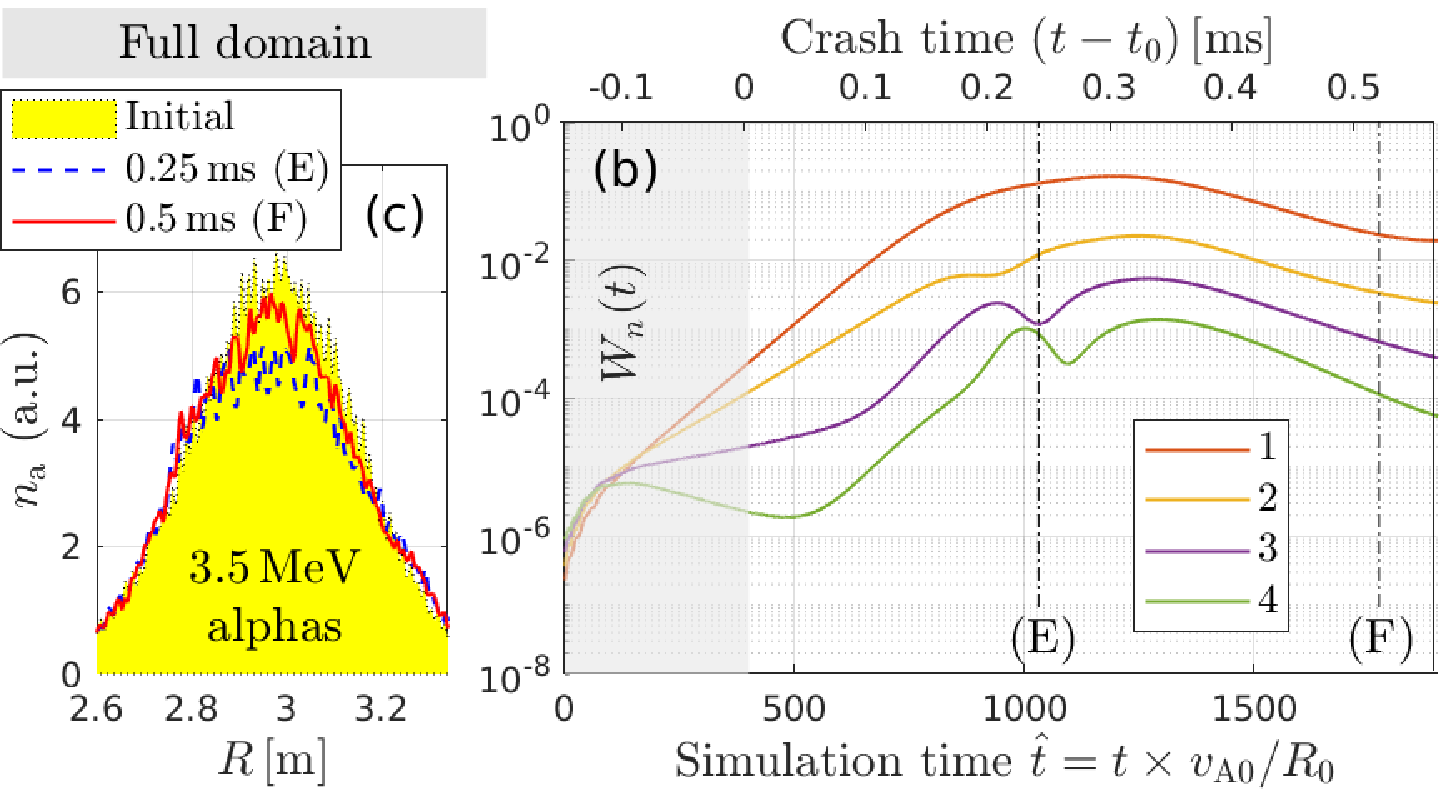}
\caption{(a) Evolution of the four dominant toroidal harmonics $n=1,2,3,4$ of the fluctuating MHD field energy $W_n(t)$ in the case with $q_0 = 0.98$, simulating only the inner 25\% of the magnetic flux space. This is a different view of the same instability as in Fig.~\protect\ref{fig:03_mega_saw}(a) of the main article. (b) Evolution of $W_n(t)$ in the simulation of the full domain for the same case as in (a). (c) Evolution of the density profile $n_{\rm a}(R)$ of $3.5\,{\rm MeV}$ alpha particles in the full-domain simulation, which confirms the result shown in Fig.~\protect\ref{fig:04_mega_kin_alpha}(e) of the main article.}
\label{fig:s03_evol_enr_alpha}%
\end{figure}

\subsection*{2. Fourier harmonics of the internal kink instability}

With $N_\zeta = 96$ grid points along the toroidal angle $\zeta$, our simulation can effectively capture toroidal Fourier harmonics $\exp(in\zeta)$ with toroidal mode numbers $-48 \leq n \leq 48$. Our simulation starts from a small initial perturbation applied to the Fourier harmonics $n=1,2,3,4$. Their evolution is shown in Fig.~\ref{fig:s03_evol_enr_alpha} in terms of the MHD fluctuation energy $W_n(t)$ as defined in Eq.~(18) of \cite{Bierwage16c}. After the initial mode structure formation process, the internal kink instability in this simulation is clearly dominated by the $n=1$ harmonic at all times.

Most results reported in the main article were obtained by simulating only the inner 25\% of the flux space and placing a non-slip boundary along the red line in Fig.~\ref{fig:01_equil}. The evolution of the MHD fluctuation energy in that simulation is shown in Fig.~\ref{fig:s03_evol_enr_alpha}(a). That simulation ran for a few $100\,\mu{\rm s}$ before macroscopic transport became visible, and we measure the crash time $t - t_0$ starting approximately from that instant. Here we have chosen $t_0 = 0.38\,{\rm ms}$, which corresponds to $\hat{t}_0 = t_0/\tau_{\rm A0} = 1000$ in units of the toroidal Alfv\'{e}n time $\tau_{\rm A0} = R_0/v_{\rm A0}$. Figure~\ref{fig:s03_evol_enr_alpha}(a) shows that the dominant harmonics are still growing exponentially at that time.

\begin{figure*}
[tbp]
\centering\vspace{-0.25cm}
\includegraphics[width=0.9\textwidth]{\figures/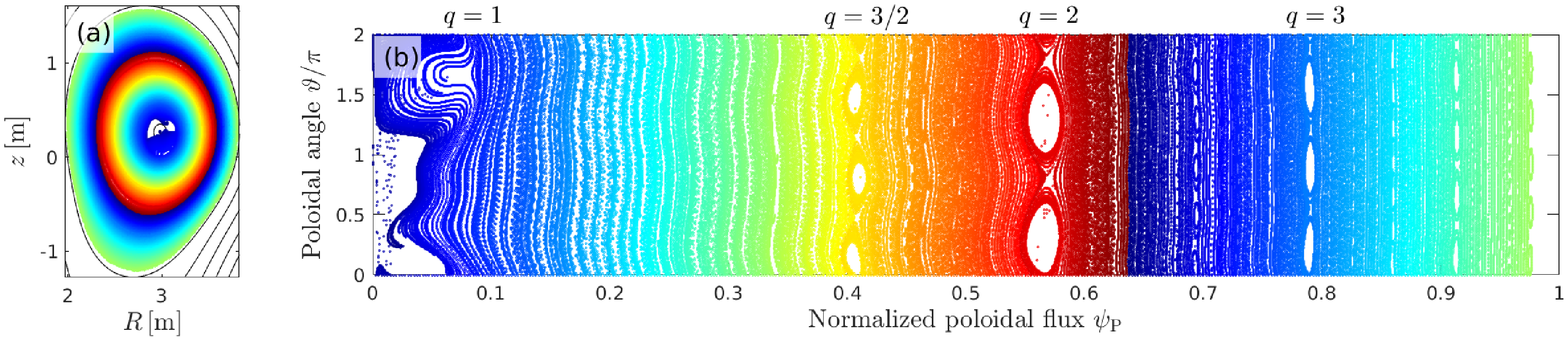}
\caption{Poincar\'{e} plots of the magnetic topology after the sawtooth crash in the full-domain simulation in Fig.~\protect\ref{fig:s03_evol_enr_alpha}(b) for snapshot (E) at $0.25\,{\rm ms}$. Different magnetic flux surfaces can be discerned by different colors. For orientation, panel (a) shows the result in cylinder coordinates $(R,z)$ at $\zeta_0 = 0$. Panel (b) shows the same data in polar coordinates $(\psi_{\rm P},\vartheta)$, where magnetic islands can be clearly seen. The normalized poloidal flux $0 \leq \psi_{\rm P} \leq 1$ is used as a minor radial coordinate ($0 =$ axis, $1 =$ edge) and is approximately proportional to $\overline{r}^2$ \protect\cite{Bierwage21b}. The locations of resonant surfaces with $q = 1,\, 3/2,\, 2/1,\, 3/1$ are indicated.}
\label{fig:s04_poin_mhd-full_islands}%
\end{figure*}

\subsection*{3. Comparison with full-domain simulation: \\ Mode evolution, fast alphas \& peripheral islands}

Simulating only the reduced domain facilitated parameter scans and convergence tests by reducing the computational effort. However, the artificial non-slip boundary located at the red line in Fig.~\ref{fig:01_equil} has a weakly stabilizing effect on the internal kink instability. Moreover, it enhances the prompt loss of a certain group of fast alpha particles that are subject to large magnetic drifts \cite{Bierwage21b}. Therefore, we performed some our our simulation in the full domain.

A comparison between panels (a) and (b) in Fig.~\ref{fig:s03_evol_enr_alpha} shows that the MHD fluctuations grow more rapidly in the full domain. Nevertheless, the sawtooth crash still has the same overall time scale of a few $100\,\mu{\rm s}$ and the $n=1$ harmonic is dominant in both simulations.

Figure~\ref{fig:s03_evol_enr_alpha}(c) shows the density profile of $3.5\,{\rm MeV}$ alphas before, during and after the crash. Apart from the larger noise (which is due to the smaller number of particles per grid cell), the result is very similar to that shown in Fig.~\ref{fig:04_mega_kin_alpha}(e) of the main article for the reduced domain. Of course, the additional boundary losses would be evident in velocity space, so our analyses of alpha particle redistribution in pitch angle and energy were performed using results of full-domain simulations (see Fig.~\ref{fig:05_pitch} of the main article, and Figs.~\ref{fig:s08_pitch} and \ref{fig:s09_enr-pitch} below).

Returning once more to the stabilizing effect of the artificial non-slip boundary: this effect can be understood by noting that the magnetic flux surfaces across the entire plasma radius are distorted in the presence of an internal kink mode. More work must be done to deform magnetic flux surfaces when the rigid boundary is closer to the $q=1$ surface, so the kink mode's growth rate is smaller in simulations with a smaller plasma.

Conversely, in the full domain, the kink can cause distortions of other low-rational resonant surfaces, such as $q = 3/2,\, 2/1,\, 3/1, ...$, and lead to the formation of magnetic islands (or enlarge existing ones) through driven magnetic reconnection. This is an actual problem in real plasmas, sometimes leading to major disruptions that terminate the discharge. For this reason, `giant' sawtooth crashes should be avoided. This was, in fact, another motivation for us to focus on cases with $q \sim 1$.

Nevertheless, even the benign sawtooth crash in our simulations with $q_0 = 0.98$ does cause magnetic reconnection at other resonant surfaces as shown in Fig.~\ref{fig:s04_poin_mhd-full_islands}. These islands are however sufficiently small and sufficiently far apart to avoid overlaps, chaos and associated loss of confinement. In any case, it must be kept in mind that Fig.~\ref{fig:s04_poin_mhd-full_islands} is the result of a resistive MHD simulation with $S = 10^6$, which may not be representative for the situation in a real and effectively collisionless tokamak plasma. The islands in Fig.~\ref{fig:s04_poin_mhd-full_islands} may be larger than they would be in reality for the same parameters.

\begin{figure}
[tbp]
\centering
\includegraphics[width=0.45\textwidth]{\figures/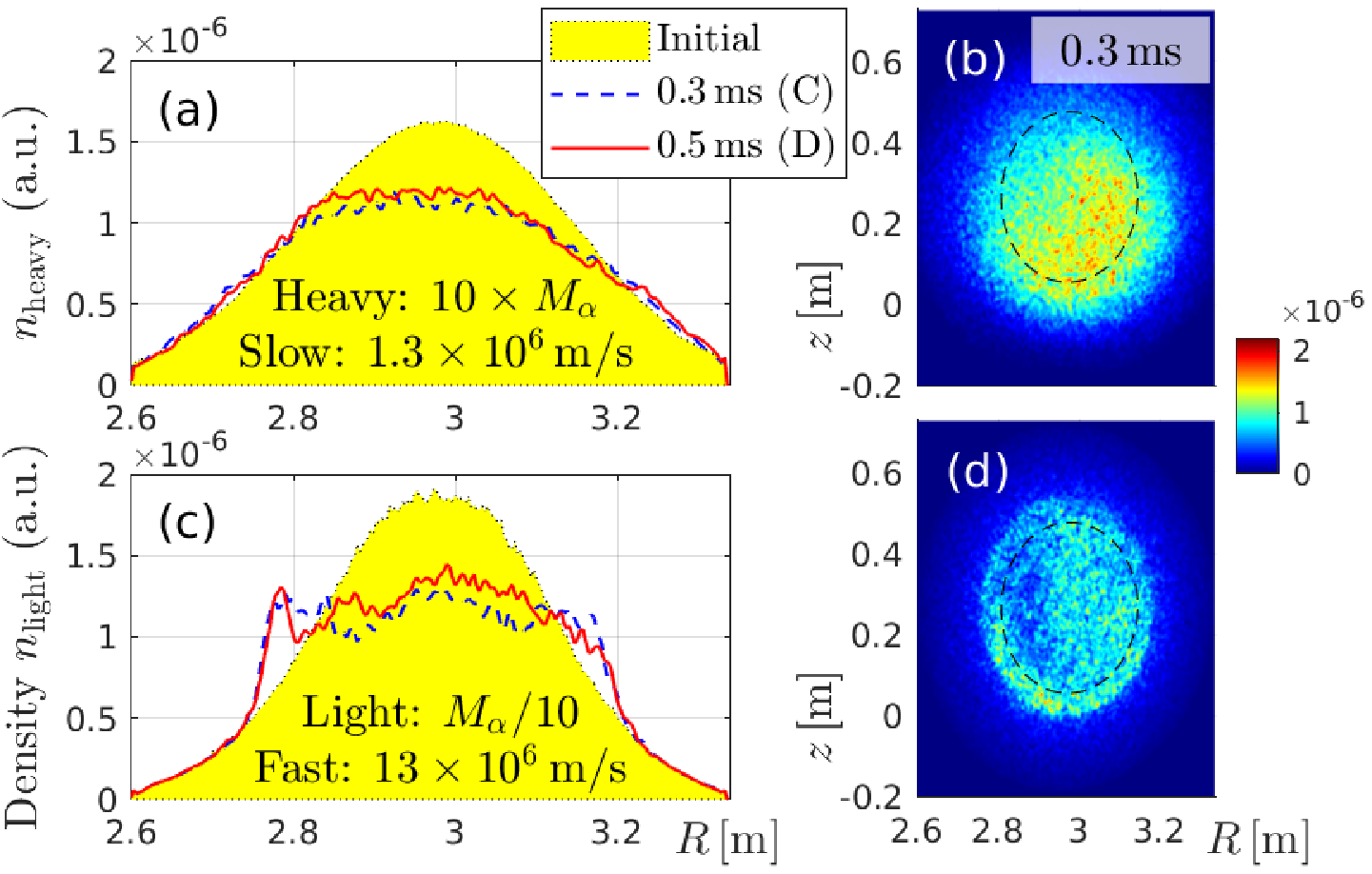}
\caption{Particle mass scan isolating the effect of magnetic drifts (top) and high particle velocity (bottom). Arranged as Fig.~\protect\ref{fig:04_mega_kin_alpha} of the main article.}
\label{fig:s05_mega_kin_10mass10}%
\end{figure}

\subsection*{4. Magnetic drift effect: I. Particle transport}

Magnetic drifts reduce the spatial corrugation of the alpha particle density field in Fig.~\ref{fig:04_mega_kin_alpha}(f) of the main article, but are not sufficient to prevent profile flattening. This is demonstrated in Fig.~\ref{fig:s05_mega_kin_10mass10}(a), which shows the redistribution of an artificial particle species with charge number $Q=2$, mass $M_{10} = 10\times M_{\rm a}$ and speed $v = 1.3\times 10^6\,{\rm m/s}$. Its magnetic drift is as large as that of fast $3.5\,{\rm MeV}$ alphas with mass $M_{\rm a}$, while the transit frequency is equal to that of slow $35\,{\rm keV}$ alphas. While the density field in Fig.~\ref{fig:s05_mega_kin_10mass10}(b) is blurred by large magnetic drifts and gyroradii as in Fig.~\ref{fig:04_mega_kin_alpha}(f), the density profile in Fig.~\ref{fig:s05_mega_kin_10mass10}(a) is subject to much stronger flattening than in Fig.~\ref{fig:04_mega_kin_alpha}(e). This must be due to the particle speed being smaller by a factor 10 since all other parameters are identical.

Figure~\ref{fig:s05_mega_kin_10mass10}(c) shows that in the opposite limit, where light particles ($M_{0.1} = M_{\rm a}/10$, $Q=2$) travel at high speed ($v = 13\times 10^6\,{\rm m/s}$), the density profile is also flattened. This, of course, is consistent with the experimentally known fact that electrons undergo radial mixing during sawtooth crashes. The results in Fig.~\ref{fig:s05_mega_kin_10mass10} thus show that the effect of magnetic drifts, which (for a given $v$) is larger for particles with smaller charge-to-mass ratio $Qe/M$, is not sufficient but necessary for preventing profile flattening.

We note that another factor that may affect our artificial light particle species in Fig.~\ref{fig:s05_mega_kin_10mass10}(c,d) is the parallel electric field $E_\parallel = {\bm E}\cdot{\bm B}/B$ (here about $1\%$ of $|{\bm E}_\perp|$) that is generated in the relaxing domain of our resistive MHD simulation. The larger charge-to-mass ratio $Qe/M$ of the light particles makes them more responsive to $E_\parallel$ than alphas and heavier ions. The consequences of this remain to be examined.

As mentioned in the Methods section of the main article, gyroaveraging had no significant effect on the results of the simulations discussed here. Nevertheless, our simulations were performed using 4-point averaging over the Larmor radius as described in \cite{Bierwage16c}, and the associated smoothing effect contributes to the form of the density fields shown in Figs.~\ref{fig:04_mega_kin_alpha} and \ref{fig:s05_mega_kin_10mass10}.

\begin{figure}
[tbp]
\centering
\includegraphics[width=0.45\textwidth]{\figures/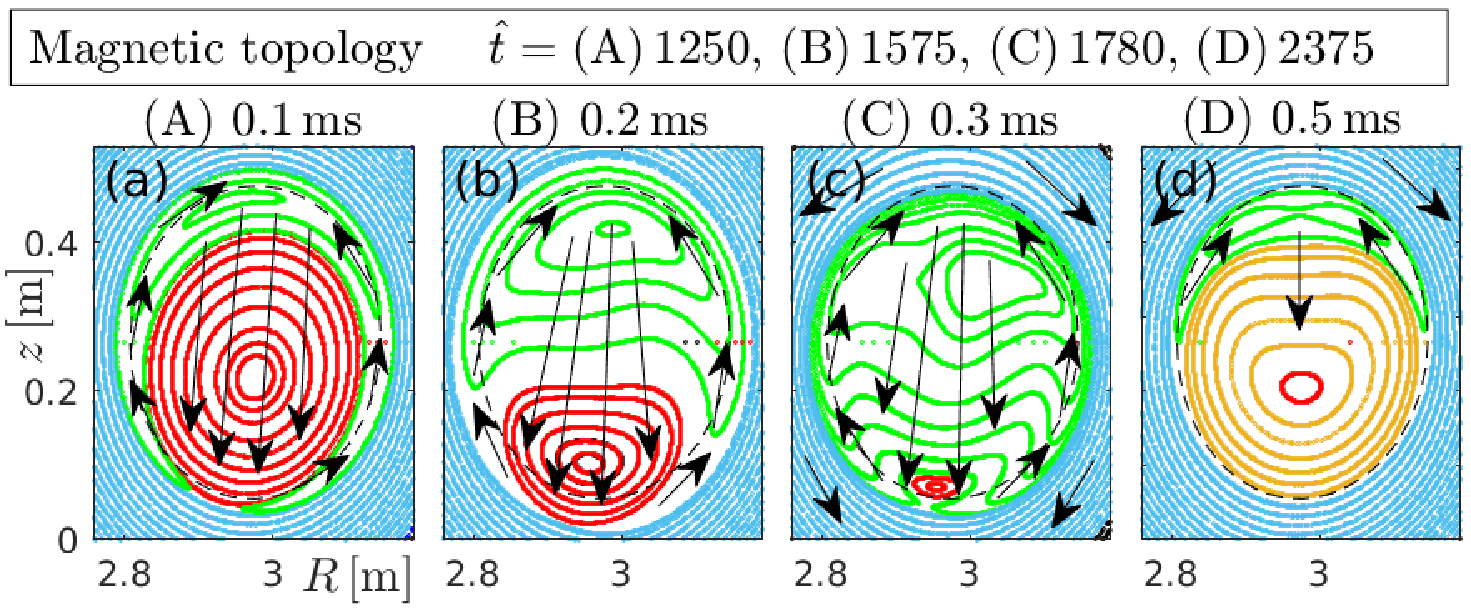}\vspace{-3.35cm} \\
\includegraphics[width=0.45\textwidth]{\figures/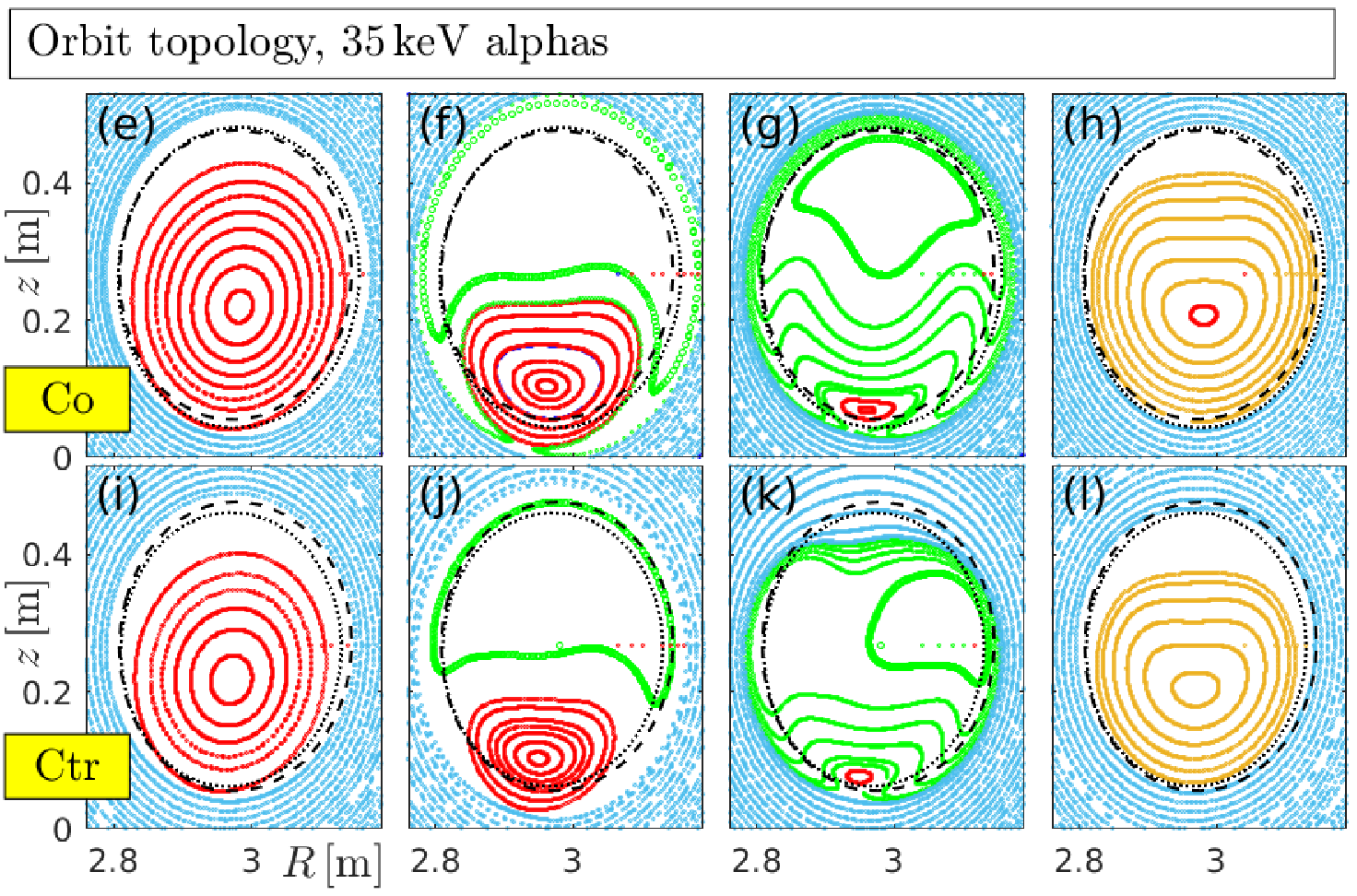}\vspace{-1.4cm} \\
\includegraphics[width=0.45\textwidth]{\figures/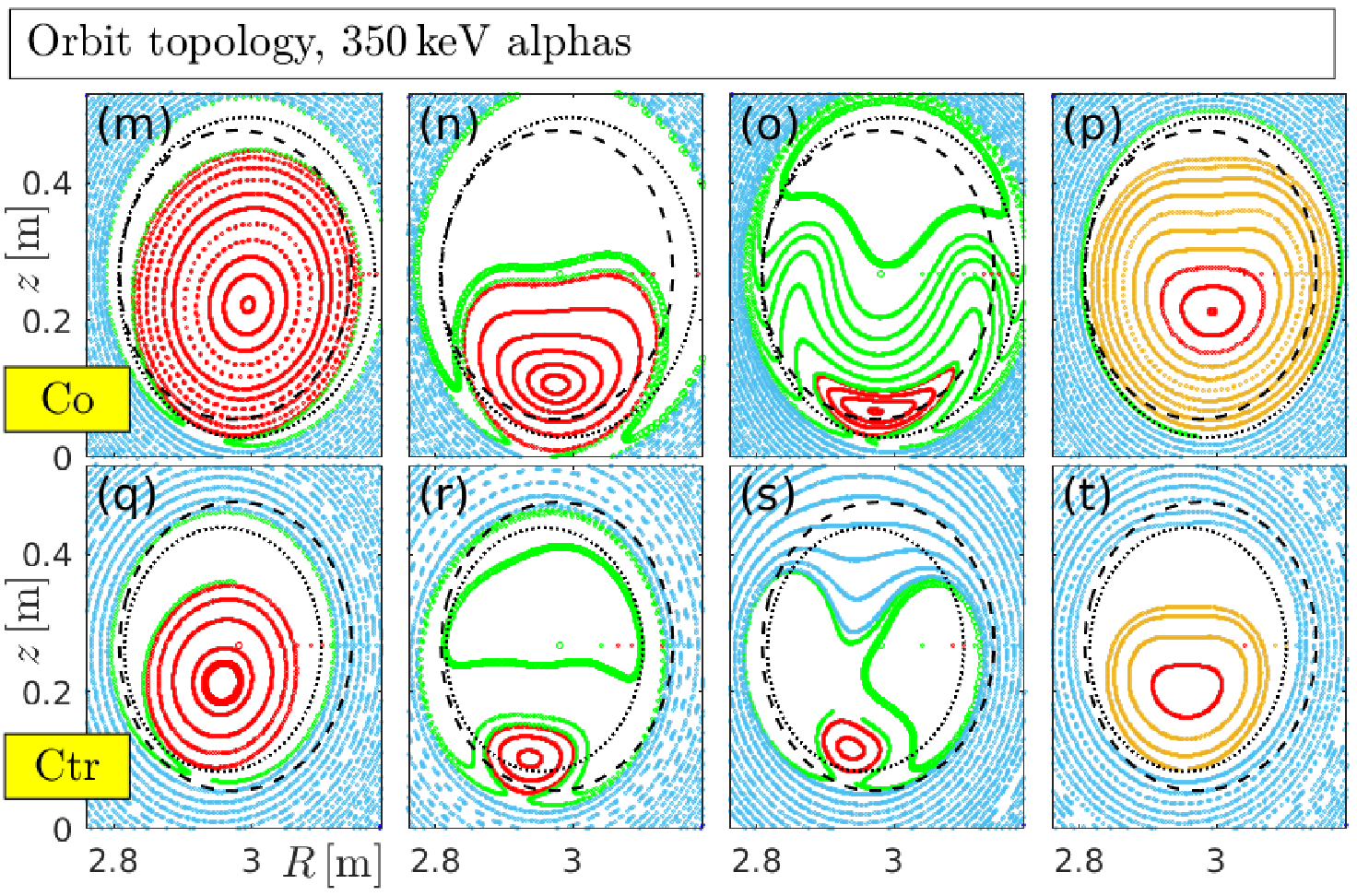}\vspace{-1.4cm} \\
\includegraphics[width=0.45\textwidth]{\figures/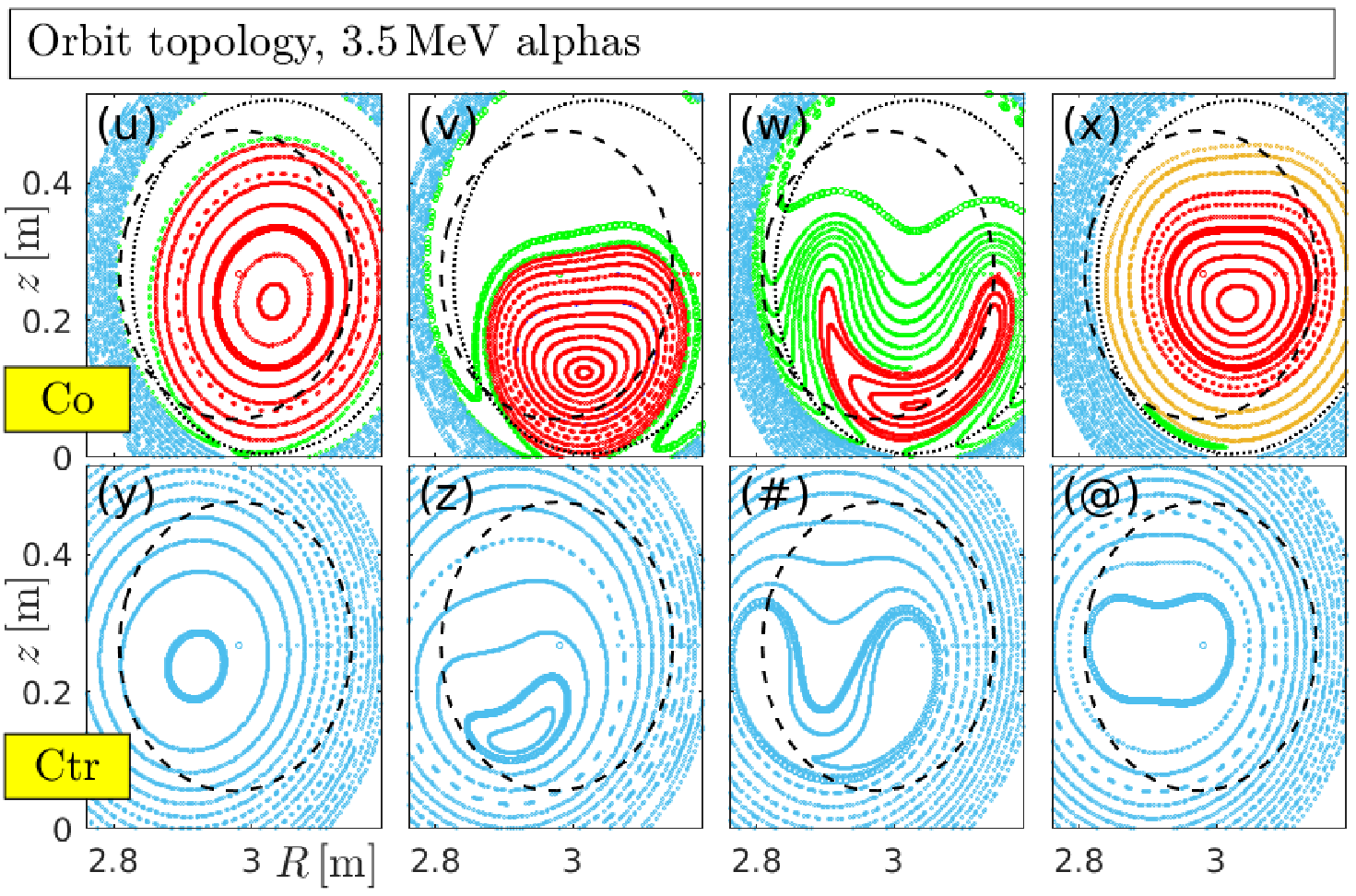}\vspace{-1.4cm}
\caption{Topology of the magnetic field (top row) and alpha particle orbits at the times of snapshots (A)--(D) indicated in Fig.~\protect\ref{fig:s03_evol_enr_alpha} (and Fig.~\protect\ref{fig:03_mega_saw}(a) of the main article). Red and orange Poincar\'{e} contours have helicities $q,h < 1$. Reconnected green islands and unreconnected blue periphery have $q,h > 1$. The initial $q = 1$ surface is drawn as a dashed circle and the dotted circles represent the initial $h = 1$ surfaces in cases where they exist. Arrows roughly indicate the direction of electric drifts. Note that the density of the field contours and orbit contours in these plots is arbitrary, so it does not by any means reflect the density of magnetic flux or orbit surfaces.}
\label{fig:s06_poin}%
\end{figure}

\subsection*{5. Magnetic drift effect: II. Orbit topology}

The Poincar\'{e} plots for the magnetic field shown in Fig.~\ref{fig:03_mega_saw}(b-d) of the main article were computed by following test particles without magnetic drifts. These plots are shown again in the top row of Fig.~\ref{fig:s06_poin}, with the addition of a fourth snapshot (D) taken near the end of the simulation at $0.5\,{\rm ms}$.

The rest of Fig.~\ref{fig:s06_poin} shows the Poincar\'{e} contours of co- and counter-passing alpha particles with energies $35\,{\rm keV}$, $350\,{\rm keV}$ and $3.5\,{\rm MeV}$. Panels (w) and (\#) already appeared in box (ii) of Fig.~\ref{fig:06_synergy} of the main article to illustrate that magnetic drifts reduce the amount of reconnection that occurs in orbit topology. The complete set of snapshots in Fig.~\ref{fig:s06_poin} confirms this statement and illustrates how the effect increases with increasing particle energy.

The orange contours appearing in snapshot (D) in the right column of Fig.~\ref{fig:s06_poin} are meant to represent regions where $q < 1$ has been restored due to the effect of the loop voltage (current source), which balances the global dissipation of the plasma current by the relatively large resistivity (small Lundquist number $S = 10^6$) in our simulations. The source term gradually restores the initial $q$ profile and would later lead to the growth of a new instability. However, the effect of the current source on the Poincar\'{e} plots is difficult to quantify, so the boundary between red and orange domains was chosen somewhat arbitrarily, based on no more than an intuitive guess.

It is interesting to note that all Poincar\'{e} plots for our default case with $q_0 = 0.98$ consist of closed contours, which means that they represent conservative Hamiltonian dynamics. No chaotic domains were seen in this case, which means that there are no significant resonance overlaps. The same is true for the case with $q_0 = 0.95$, whereas signatures of chaos near the mixing radius can be seen in the non-monotonic case with $q_0 = 0.87$ that will be discussed in Section 7 below (Figs.~\ref{fig:s11_scan2_poin} and \ref{fig:s12_scan4_poin}).

\begin{figure}
[tbp]
\centering
\includegraphics[width=0.45\textwidth]{\figures/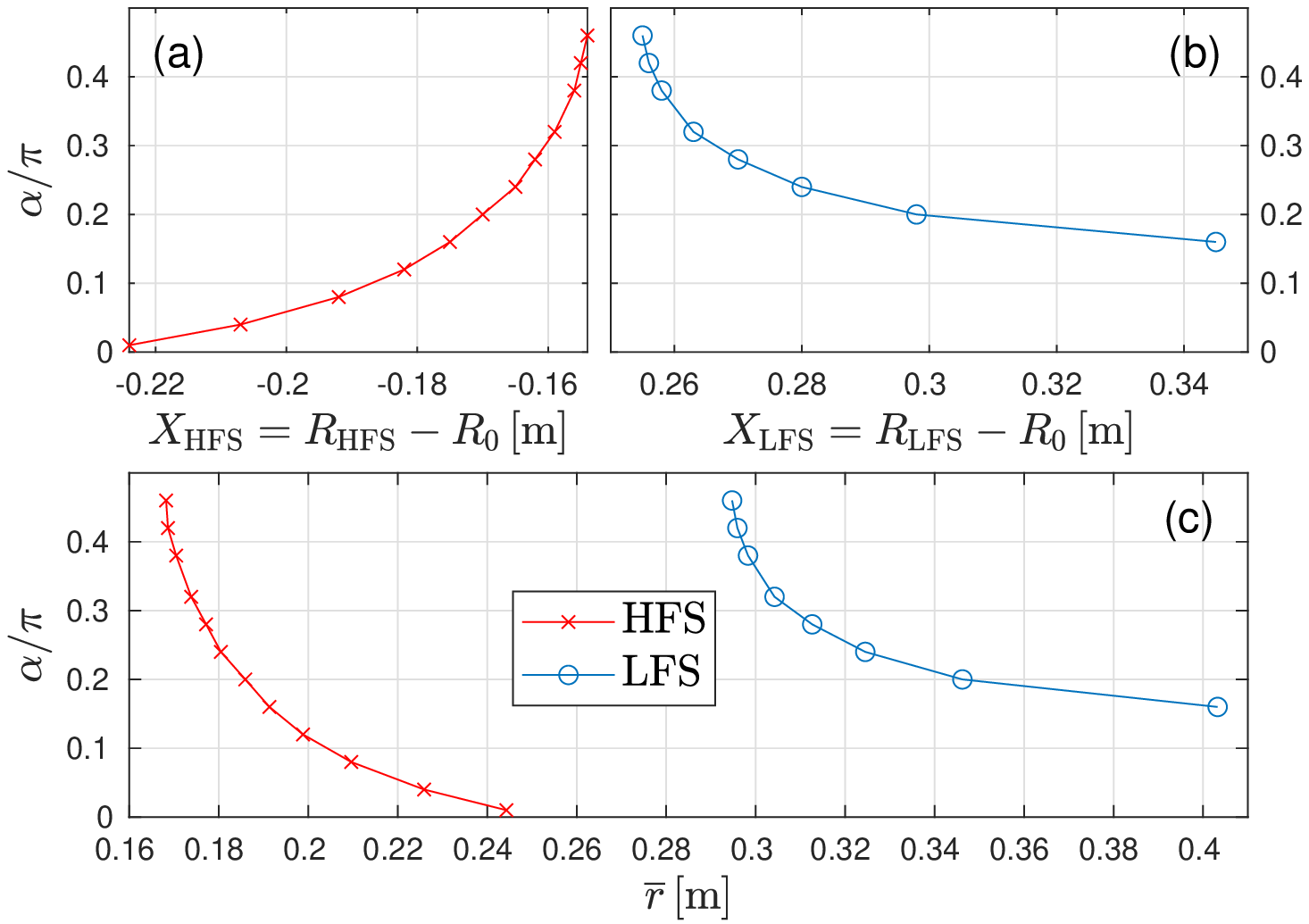}\vspace{-0.15cm}
\caption{High-field-side (HFS, $R < R_0$) and low-field-side (LFS, $R > R_0$) radii of the $h = 1$ resonance on the midplane ($z_0 = 0.26$) for co-passing $3.5\,{\rm MeV}$ alpha orbits in the default case with $q_0 = 0.98$. The $h = 1$ resonance is absent for counter-passing alphas with $3.5\,{\rm MeV}$ in this case.}
\label{fig:s07_jet_3500keV_resonance_X-r-pitch}%
\end{figure}

Our Poincar\'{e} analyses were focused on particles in the deeply passing domain; that is, on pitch angles near $\pm 90$ degrees: $v_\parallel/v = \sin(\pm 0.48\pi) \approx \pm 1$. For completeness, Fig.~\ref{fig:s07_jet_3500keV_resonance_X-r-pitch} shows the pitch dependence of the $h=1$ resonance radius for alpha particles with kinetic energy $K = 3.5\,{\rm MeV}$. Both the inner (high-field-side) and outer (low-field-side) radius is shown. The respective points closest to the magnetic axis --- namely, $X_{\rm HFS} = -0.154\,{\rm m}$, $\overline{r}_{\rm HFS} = 0.168\,{\rm m}$ and $X_{\rm LFS} = 0.255\,{\rm m}$, $\overline{r}_{\rm LFS} = 0.288\,{\rm m}$ --- correspond to the points where the red curve in Fig.~\ref{fig:07_helicity} crosses 1. One can see in Fig.~\ref{fig:s07_jet_3500keV_resonance_X-r-pitch} that the resonant radius quickly increases as the pitch is reduced towards the trapped-passing boundary, which is located near $\alpha \approx 0.15\pi$. This means that internal kink mode has a smaller influence on particles with smaller pitch angle, which is precisely what we observe in the simulations as can be ascertained in Fig.~\ref{fig:05_pitch}(e,f). Trapped particles and counter-passing do not resonate at all with the internal kink in this configuration ($q_0 = 0.98$).

\begin{figure}
[tbp]
\centering\vspace{-0.25cm}
\includegraphics[width=0.45\textwidth]{\figures/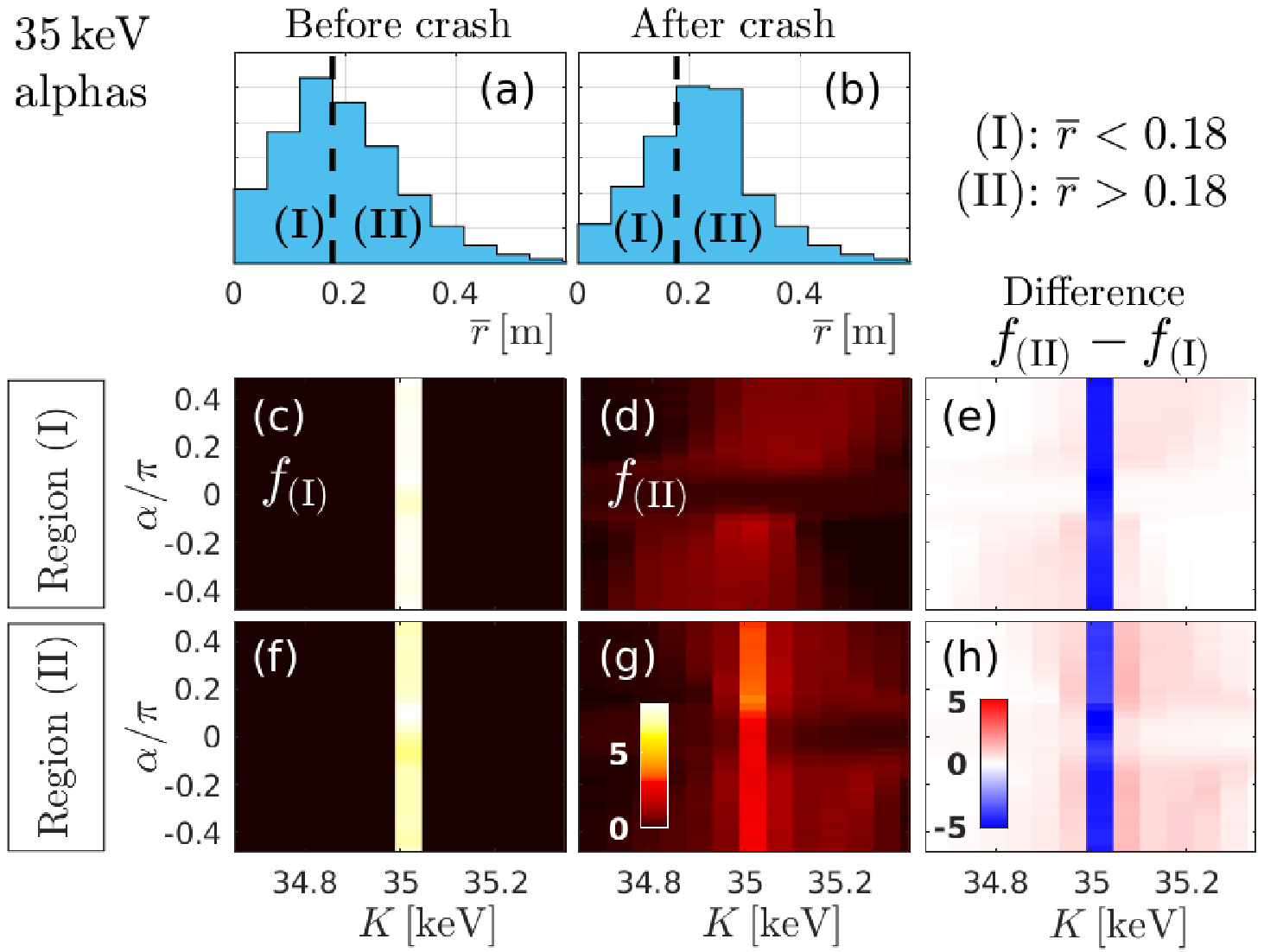} \vspace{0.2cm} \\
\includegraphics[width=0.45\textwidth]{\figures/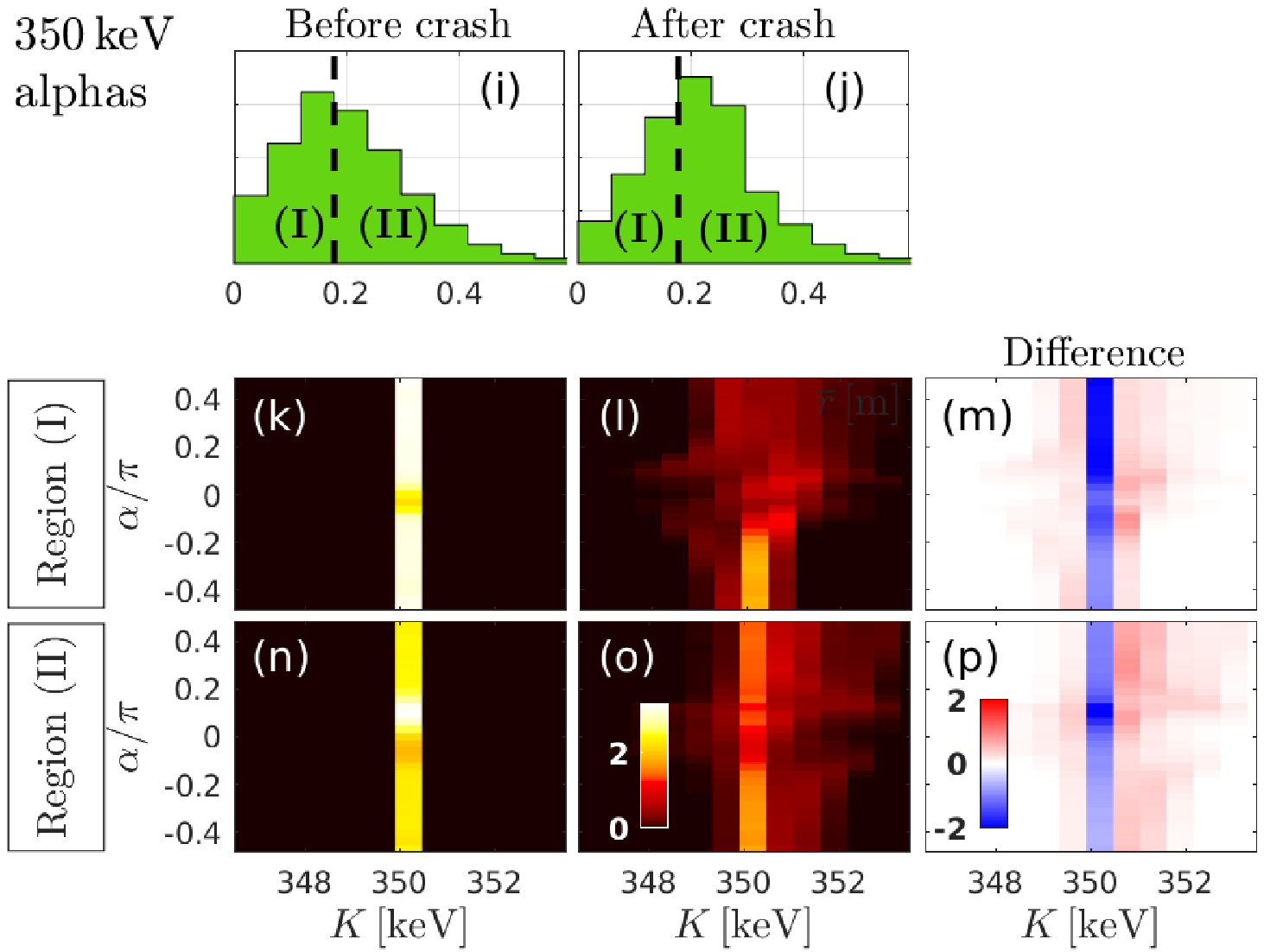} \vspace{0.2cm} \\
\includegraphics[width=0.45\textwidth]{\figures/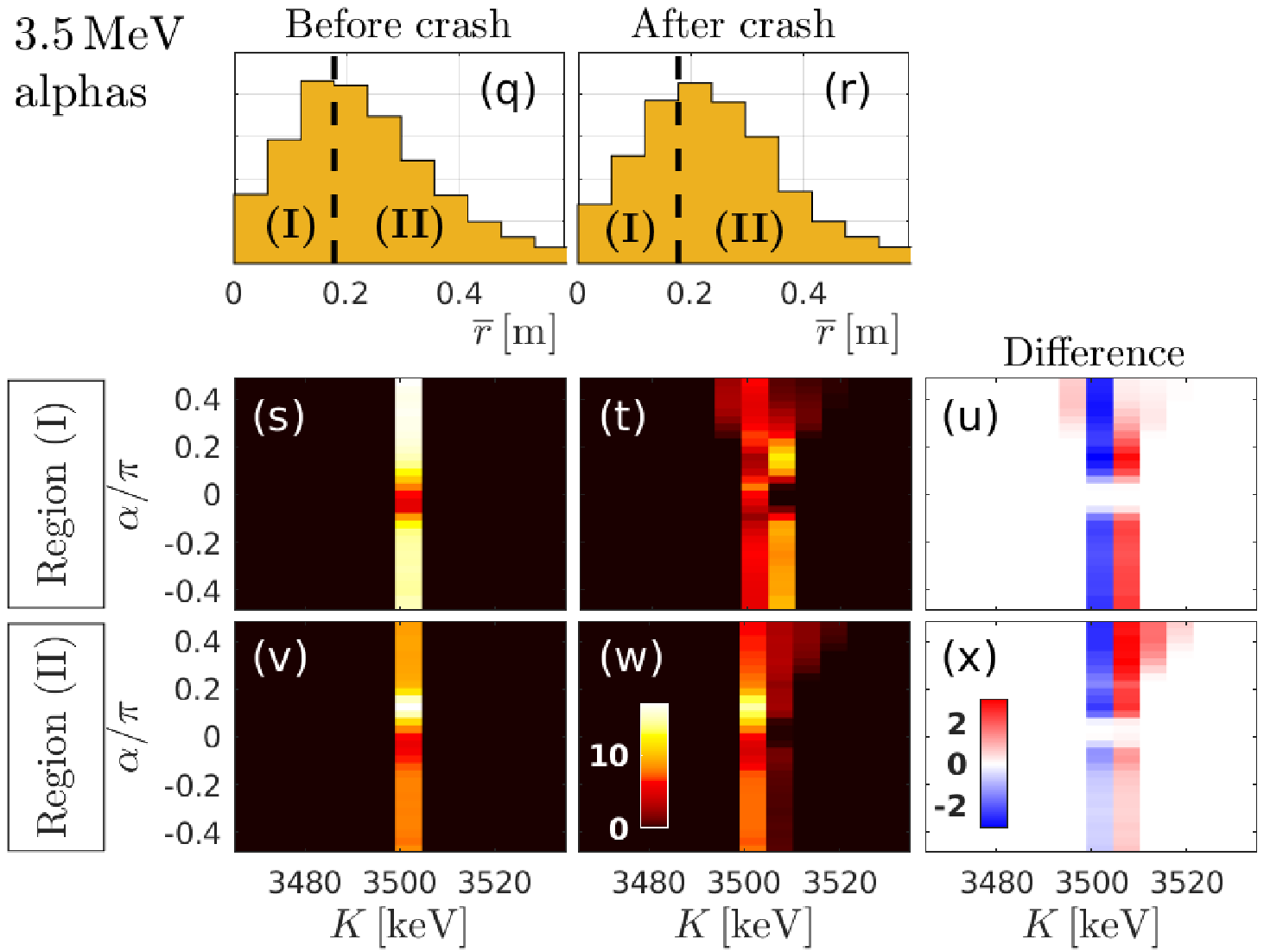}
\caption{Evolution of the alpha particle velocity distribution with initial energies $35\,{\rm keV}$ (a-h), $350\,{\rm keV}$ (i-p) and $3.5\,{\rm MeV}$ (q-x). The full domain is simulated in order to capture also large orbits at small values of $\alpha$, which pass both through the plasma core and the periphery \protect\cite{Bierwage21b}. The left column shows the initial state before the crash, the central column shows the post-crash state, and the difference is shown on the right. Panels (a,b), (i,j) and (q,r) show histograms of particle numbers as functions of the mean minor radius $\overline{r}$. The vertical dashed line at $\overline{r} = 0.18\,{\rm m}$ roughly corresponds to the $q=1$ surface $\overline{r}_1$. The inner region $\overline{r} \lesssim \overline{r}_1$ and the outer region $\overline{r} \gtrsim \overline{r}_1$ are labeled (I) and (II), respectively. The pitch angle distributions for region (I) are shown in (c-e), (k-m) and (s-u), and those for region (II) are shown in (f-h), (n-p) and (v-x).}
\label{fig:s08_pitch}%
\end{figure}

\subsection*{6. Redistribution in velocity space}

In Fig.~\ref{fig:05_pitch} of the main article, we showed how particles at different pitch angles are redistributed by the sawtooth crash in Fig.~\ref{fig:03_mega_saw} for our default case with $q_0 = 0.98$. For that purpose, we measured the particle populations in an inner region $\overline{r} < 0.18\,{\rm m}$ and an outer region $\overline{r} > 0.18\,{\rm m}$. Figure~\ref{fig:s08_pitch} provides a more detailed view of the data, showing also how the sawtooth affects the kinetic energy $K$ of the particles, which had been initialized with a single value $35\,{\rm keV}$ (a-h), $350\,{\rm keV}$ (i-p) or $3.5\,{\rm MeV}$ (q-x).

In the region $0 \leq \overline{r} \leq 1.08\,{\rm m}$ (or $0 \leq \overline{r}/\left<a\right> \leq 0.9$ in normalized units), we divided the minor radial axis into 18 cells of size $\Delta\overline{r} = 0.06\,{\rm m}$. By counting the number of particles in each radial cell, before and after the sawtooth crash, we obtained the histograms shown in panels (a,b), (i,j) and (q,r) of Fig.~\ref{fig:s08_pitch}. We emphasize that these are histograms, not densities, so that that the number of particles in each cell decreases towards the axis since the cell volume scales like $\Delta V \propto \overline{r} R \Delta\overline{r}$. The reason for using histograms instead of densities for the radial direction is that histograms show the amount of particle transport more clearly: the reduction seen in one region corresponds to the gain in another, which makes plots like those in Fig.~\ref{fig:05_pitch} quantitatively intuitive.

For each cell $\overline{r}_i$ ($i = 1,...,18$), we recorded the velocity distribution $f_i(K,\alpha)$ in the form of a true density function, which is a histogram divided by the velocity space Jacobian $\J_{K\alpha} \propto 2\pi v_\perp$ (cf.~Eq.~(A.12) of \cite{Bierwage21b}). Their integration over an inner region (I) $\overline{r} < 0.18\,{\rm m}$ and an outer region (II) $\overline{r} > 0.18\,{\rm m}$ yielded the velocity distributions $f_{\rm (I)}(K,\alpha)$ and $f_{\rm (II)}(K,\alpha)$ as well as the difference $f_{\rm (II)} - f_{\rm (I)}$ shown in panels (c-h), (k-p) and (s-x) of Fig.~\ref{fig:s08_pitch}. Here, the density function $f(K,\alpha)$ was preferred over the histogram $H(K,\alpha) = \J_{K\alpha} f(K,\alpha)$ because an isotropic distribution becomes a constant along the pitch angle $\alpha$ when shown in terms of $f(K,\alpha)$ as in Fig.~\ref{fig:s08_pitch}(c).

The boundary between regions (I) and (II) was put close to the initial $q=1$ radius $\overline{r}_1 \approx 0.19\,{\rm m}$, since the latter corresponds roughly to the inversion radius of a sawtooth crash. Although neither the minor radius $\overline{r}$ nor the pitch angle $\alpha$ of an alpha particle is conserved (even during its unperturbed motion) in a tokamak, the resulting blurring of the measured distribution is acceptable at the level of detail that is of interest here, so we keep it simple and do not use orbit-based coordinates \cite{Bierwage14b}.

The main message that we wish to convey with Fig.~\ref{fig:s08_pitch} is as follows. The majority of supra-thermal alpha particles tend to be accelerated by the kink mode, which means that they have a stabilizing influence on the mode. Figure~\ref{fig:s08_pitch} shows this effect most clearly for the co-passing particles. Our simulations also confirm the well-known stabilizing influence of trapped particles \cite{Porcelli91, Porcelli92, Porcelli94}, although it is not clearly visible in Fig.~\ref{fig:s08_pitch} because the trapped particles in the domain $-0.15\pi \lesssim \alpha \lesssim 0.15\pi$ are spread out in energy and many of them even exit the narrow energy window shown, so they are no longer visible after the crash. (Their weights are however included in the energy-integrated plots of Fig.~\ref{fig:05_pitch} of the main article.)

Some exceptions can also be seen and may be explained in terms of resonances. For instance, Fig.~\ref{fig:s08_pitch}(u) shows that some of the co-passing $3.5\,{\rm MeV}$ alphas around $\alpha \gtrsim +0.3\pi$ in region (I) are decelerated, which means that they drive the internal kink mode. This is consistent with the fact that we have initialized the simulation with a destabilizing density profile and that a significant portion of the $h = 1$ resonance of these particles lies within the $q = 1$ radius $\overline{r}_1 \approx 0.19$, as one can see from the values of $\overline{r}_{\rm HFS}$ in Fig.~\ref{fig:s07_jet_3500keV_resonance_X-r-pitch}(c).

Figure~\ref{fig:s08_pitch}(e) seems to indicate that slow alphas with $35\,{\rm keV}$ exert a net driving force in the counter-passing domain and a net damping force in the co-passing domain. However, the effect is subtle and one should exercise care when trying to interpret this observation because these particles are subject to violent transport: as we have seen in  Fig.~\ref{fig:04_mega_kin_alpha}(b), these particles move in and out and around the relaxation domain, so they may transiently resonate with the mode at one time or another whenever their orbit helicity matches that of the mode ($h = 1$). We have not made any attempt to untangle this process in detail.

\begin{figure}
[tbp]
\centering\vspace{-0.25cm}
\includegraphics[width=0.45\textwidth]{\figures/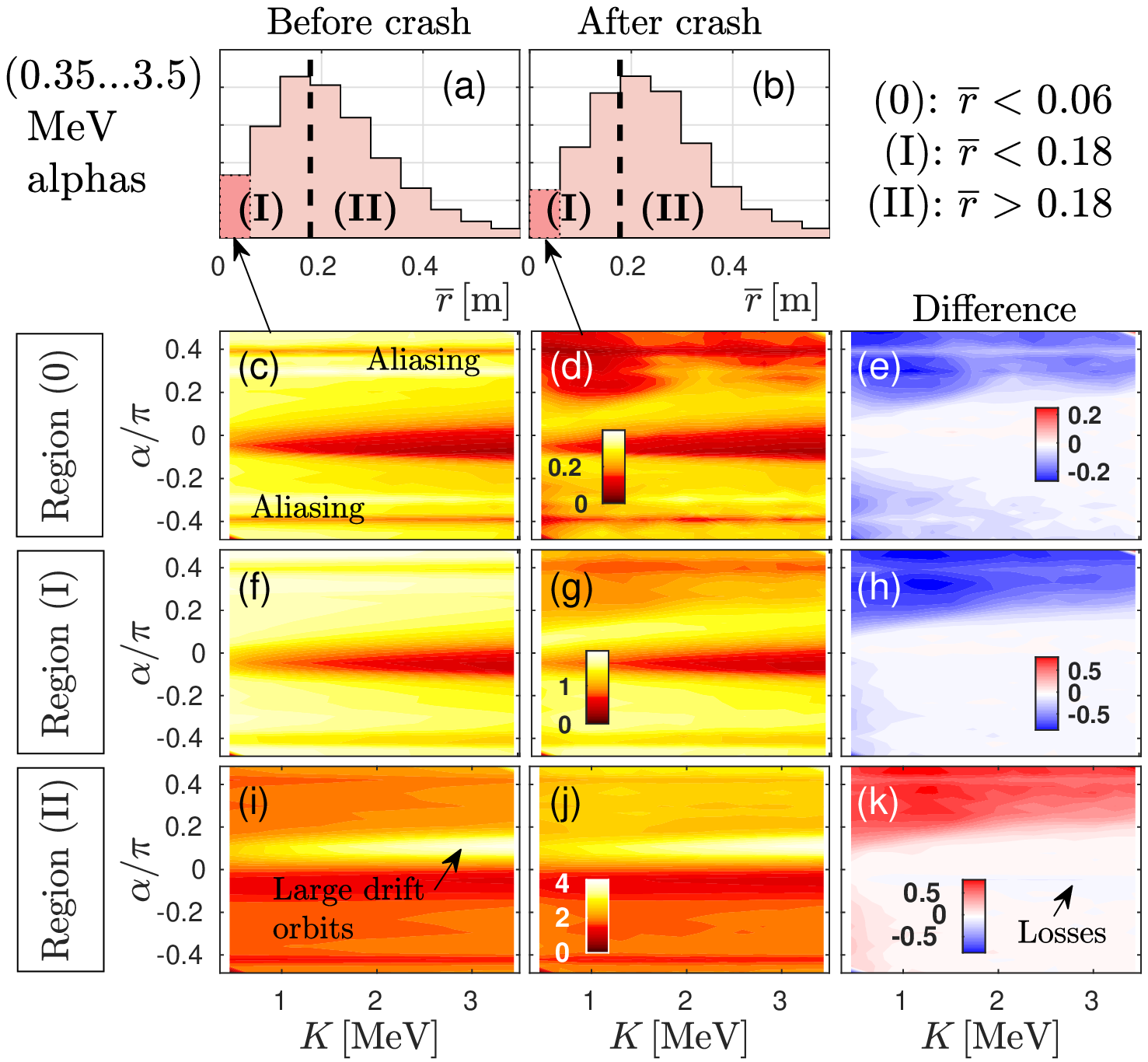}
\caption{Evolution of the alpha particle velocity distribution in a simulation initialized with a uniform population of particles in the energy range $0.35\,{\rm MeV} \leq K \leq 3.5\,{\rm MeV}$ \protect\cite{Bierwage21b}. Arranged like Fig.~\protect\ref{fig:s08_pitch}(a-h), with the addition of a third integration region $0 < \overline{r} < 0.06\,{\rm m}$ labeled `Region (0)', results for which are shown in panels (c)-(e). Note that the horizontal stripes around $\alpha \approx \pm (0.3...0.4)\pi$ are aliasing artifacts that should be ignored (cf.~Section 8.3 in \protect\cite{Bierwage21b}). This simulation was performed in the full domain, and only a vanishingly small amount of boundary losses is visible in panel (k).\vspace{-0.15cm}}
\label{fig:s09_enr-pitch}%
\end{figure}

Finally, it is interesting to inspect how smooth or sharp the energy threshold for the confinement of passing alpha particles is. For this purpose, Fig.~\ref{fig:s09_enr-pitch} shows the evolution of the velocity distribution $f(K,\alpha)$ in a simulation initialized with alpha particles distributed uniformly in the energy range $0.35\,{\rm MeV} \leq K \leq 3.5\,{\rm MeV}$.

In the domain of co-passing particles ($\alpha > 0$), the sawtooth-induced transport can be seen to vary gradually. There is no sharp threshold in the energy window examined here. In the case of counter-passing particles ($\alpha < 0$), panels (h) and (k) of Fig.~\ref{fig:s09_enr-pitch} seem to indicate that there is a transport threshold near $0.5\,{\rm MeV}$. However, as we have also noted in the main article, the observed amount of co-/counter-passing asymmetry is linked to the choice of the boundary between the `inner' and `outer' integration regions. To illustrate this, we have added in Fig.~\ref{fig:s09_enr-pitch}(c-e) the results for a smaller `inner' region $\overline{r} < 0.06\,{\rm m}$ labeled `Region (0)'. Figure~\ref{fig:s09_enr-pitch}(e) shows clearly that the transport of counter-passing alphas out of the this small inner region (0) increases gradually across an energy range of several MeV, beginning with large negative pitch angles ($\alpha \approx -\pi/2$) and then broadening towards the trapped-passing boundary ($\alpha \approx -0.15\pi$).

\begin{figure}
[tbp]
\centering\vspace{-0.25cm}
\includegraphics[width=0.45\textwidth]{\figures/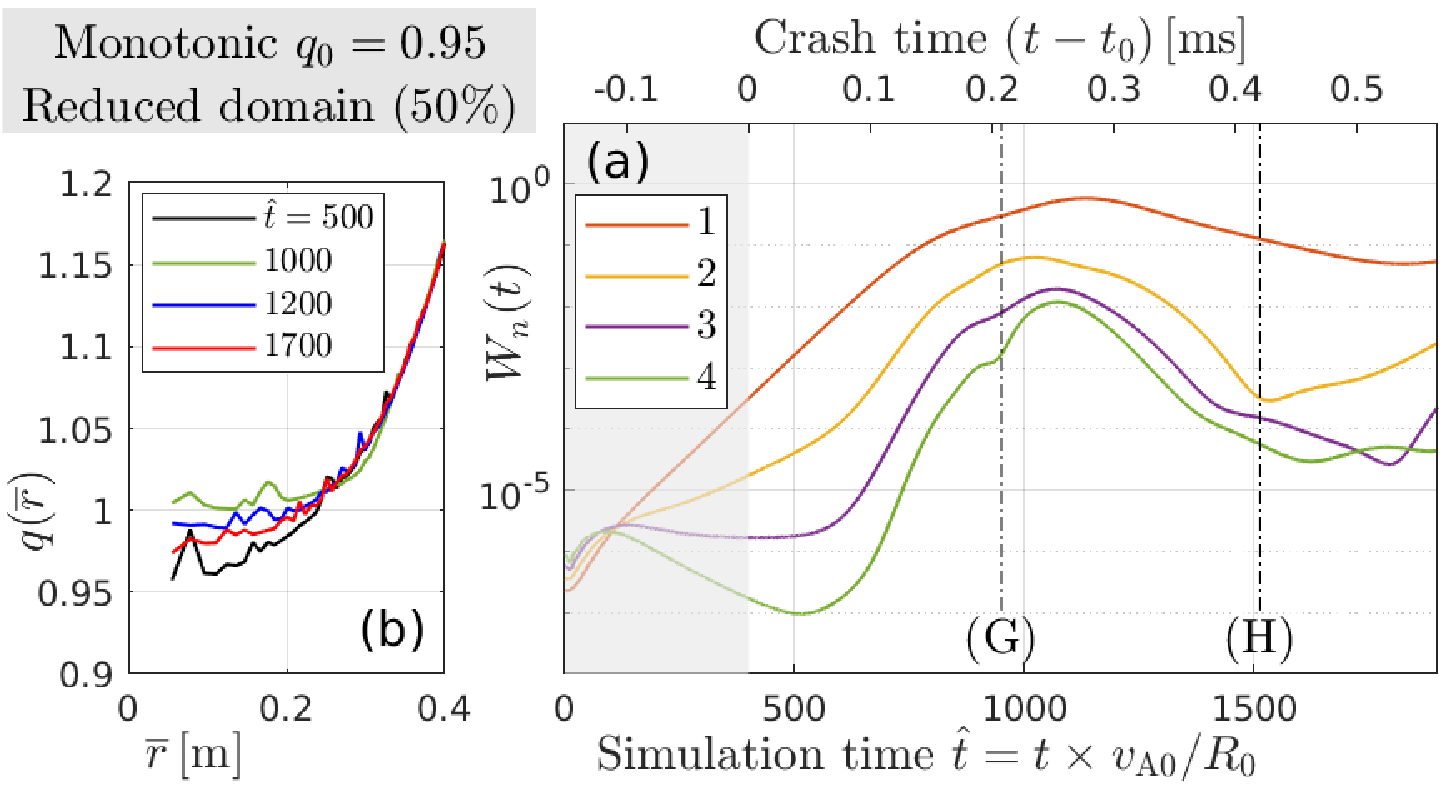} \vspace{0.2cm} \\
\includegraphics[width=0.45\textwidth]{\figures/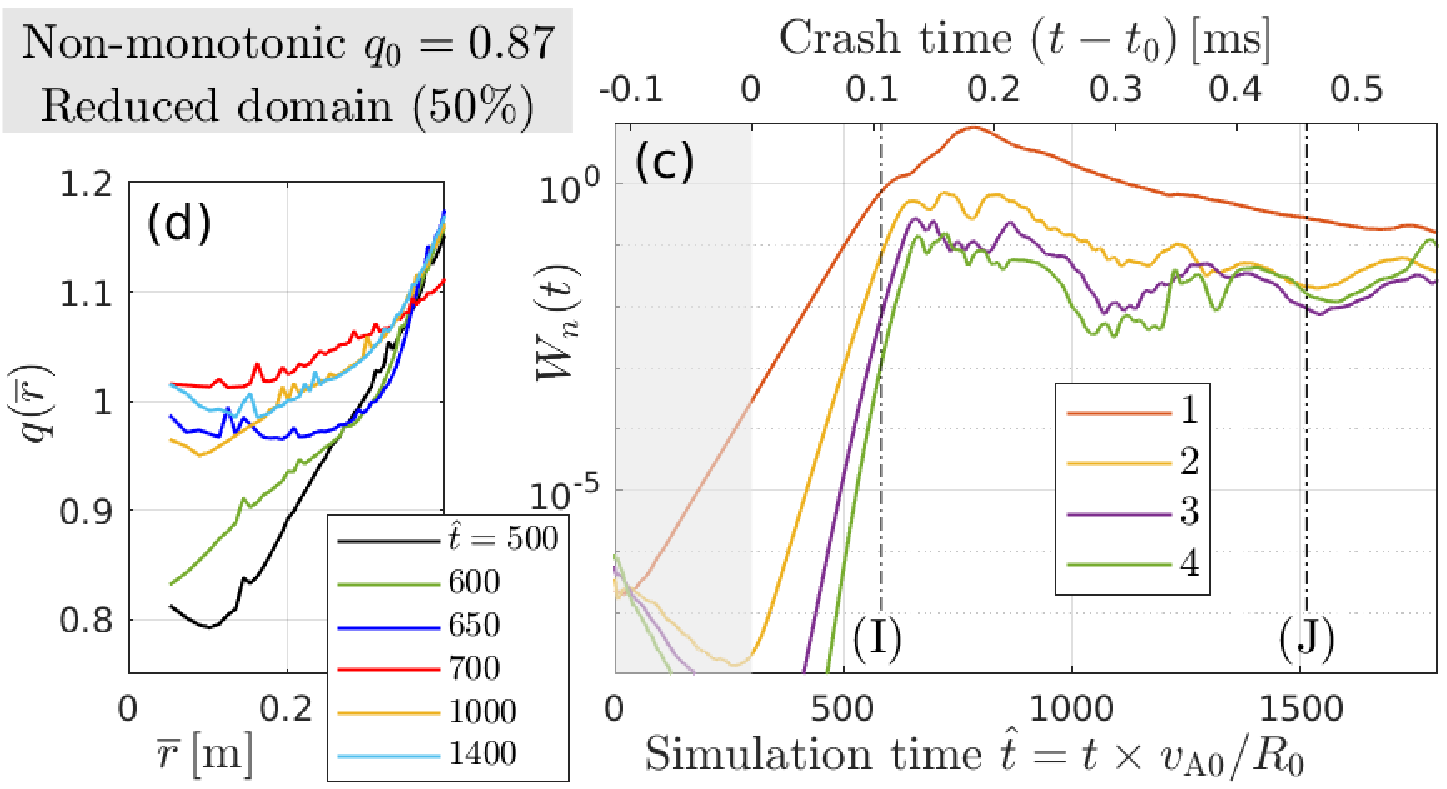}
\caption{Comparison of the evolution of (a,c) the MHD fluctuation energy $W_n(t)$ and (b,d) the $q$ profile in the cases initialized with $q_0 = 0.95$ and $q_0 = 0.87$ (cf.~Fig.~\protect\ref{fig:s02_prof_qsafe}). Panels (a) and (c) are arranged as Fig.~\protect\ref{fig:s03_evol_enr_alpha}(a,b). The labels (G)-(J) indicate the snapshot times for which alpha particle density profiles are plotted in Figs.~\protect\ref{fig:s14_scan2_kin_alpha} and \protect\ref{fig:s15_scan4_kin_alpha} below. The $q$ profiles in (b,d) represent field helicities averaged over the initial (unperturbed) magnetic surfaces.\vspace{-0.15cm}}
\label{fig:s10_evol_enr-q_scan24}%
\end{figure}

\begin{figure}
[tbp]
\centering
\includegraphics[width=0.45\textwidth]{\figures/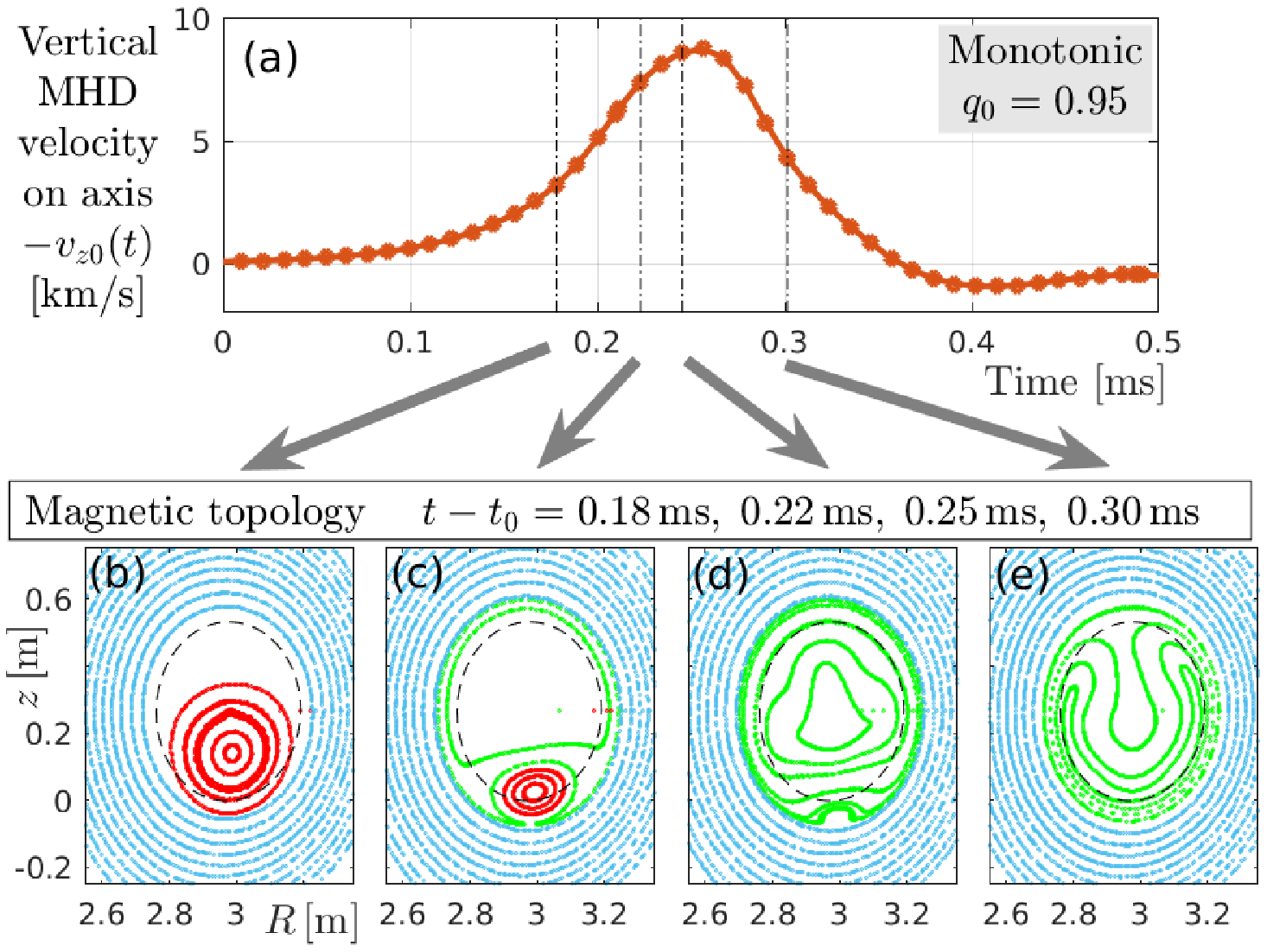}\vspace{0cm} \\
\includegraphics[width=0.45\textwidth]{\figures/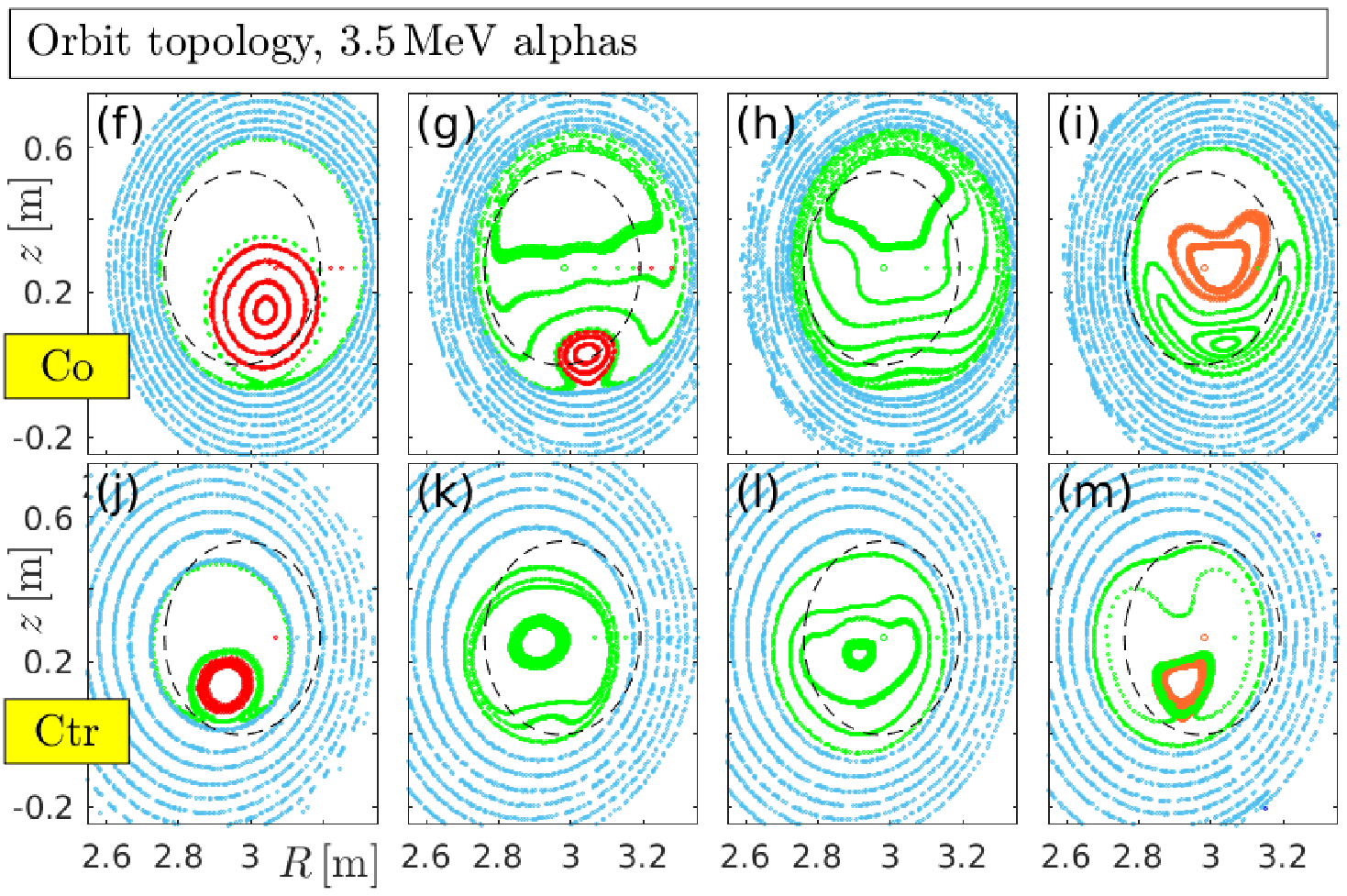}\vspace{-1.1cm} \\
\includegraphics[width=0.45\textwidth]{\figures/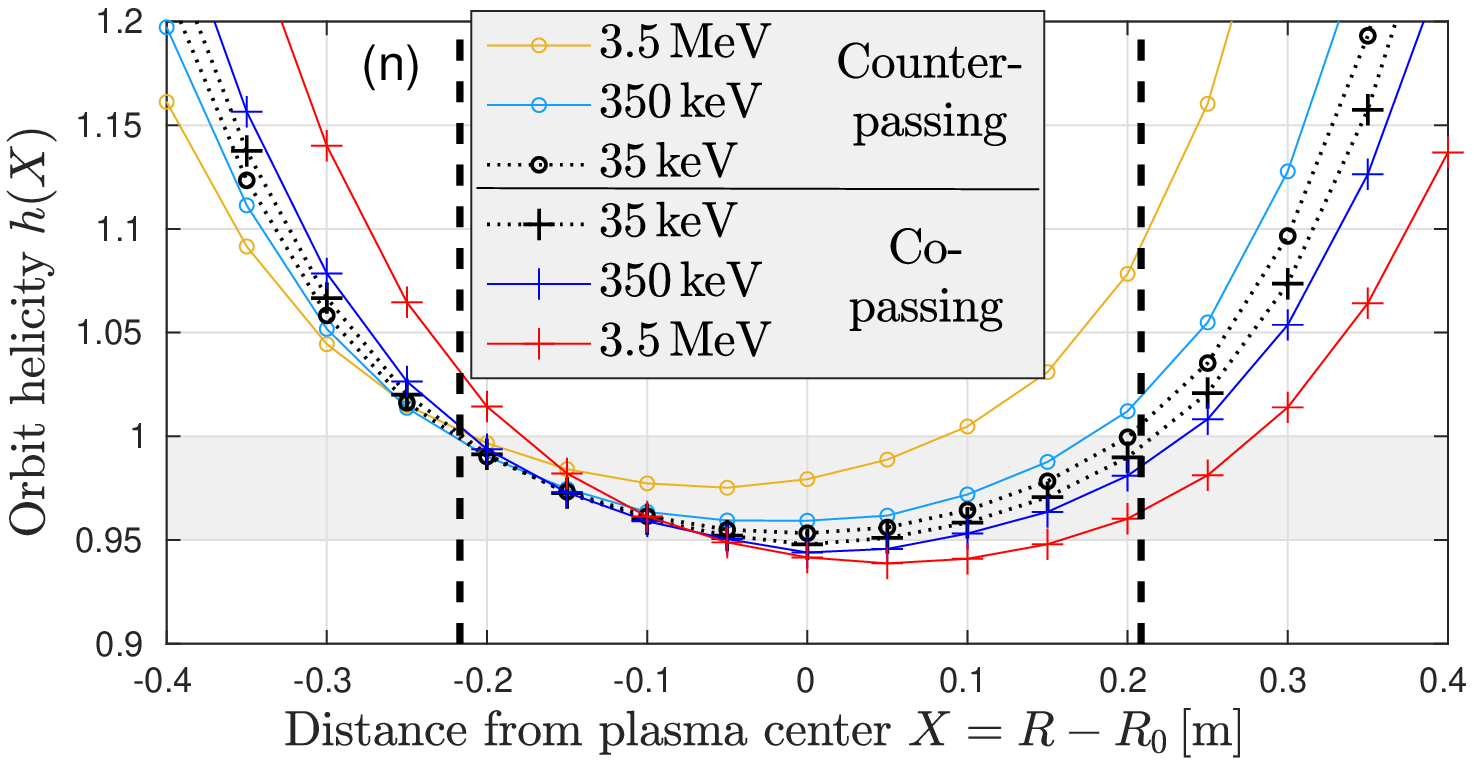}
\caption{Overview of the sawtooth crash dynamics in the case with monotonic $q$ profile and central value $q_0 = 0.95$. (a) Evolution of the vertical MHD velocity $v_{z0}(t)$ at the magnetic axis as in Fig.~\protect\ref{fig:03_mega_saw}(a). Vertical dash-dotted lines mark the times for which Poincar\'{e} plots are shown in the panels below (arrows). (b-m) Poincar\'{e} plots showing the topology of the magnetic field and alpha particle orbits with energy $K = 3.5\,{\rm MeV}$ and pitch angles $\alpha = \pm0.48\pi$. Arranged similarly to Fig.~\protect\ref{fig:s06_poin}. (n) Orbit helicity profiles $h(X)$. Arranged as in Fig.~\protect\ref{fig:07_helicity}.}
\label{fig:s11_scan2_poin}%
\end{figure}

\begin{figure}
[tbp]
\centering
\includegraphics[width=0.45\textwidth]{\figures/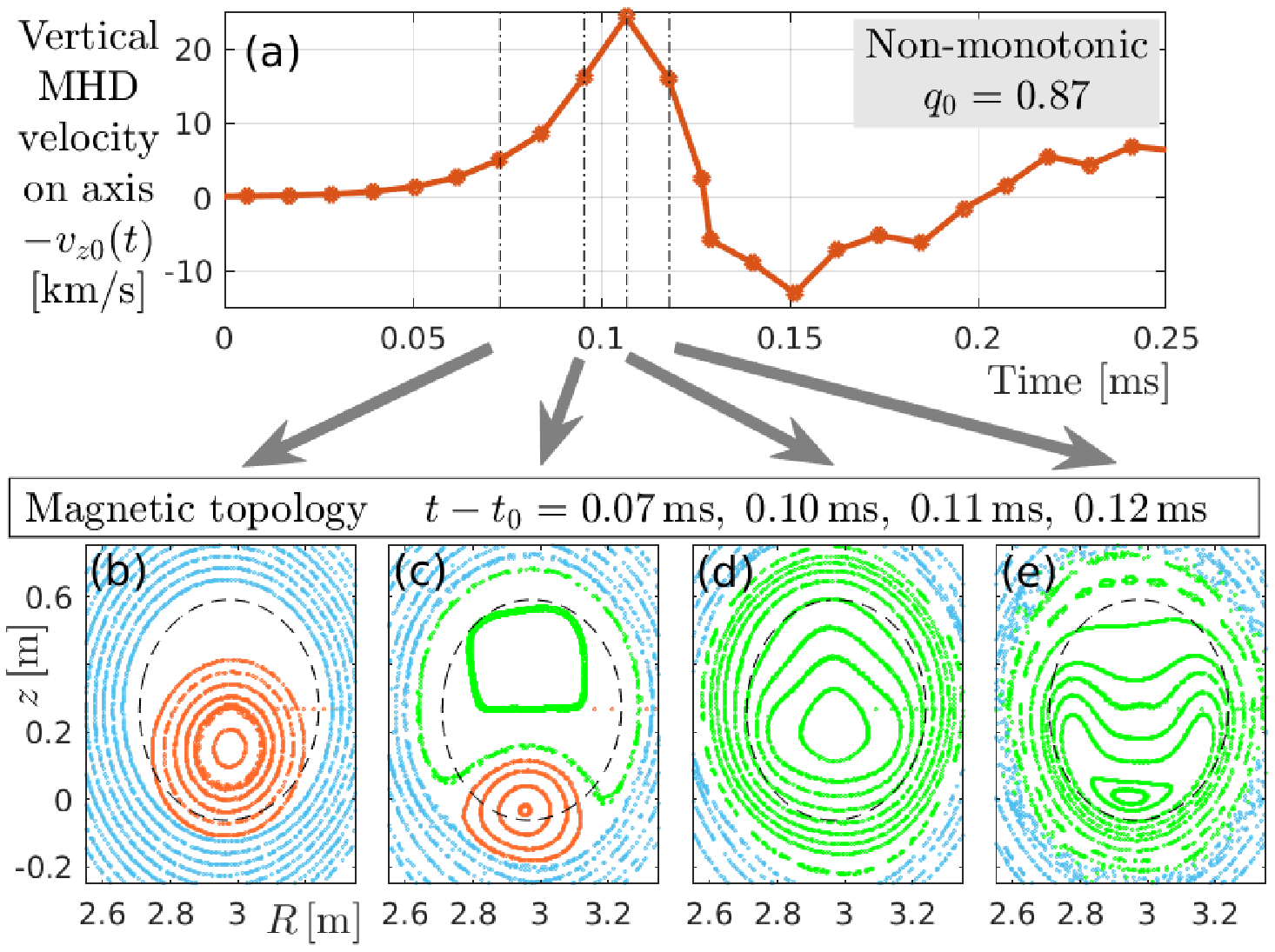}\vspace{0cm} \\
\includegraphics[width=0.45\textwidth]{\figures/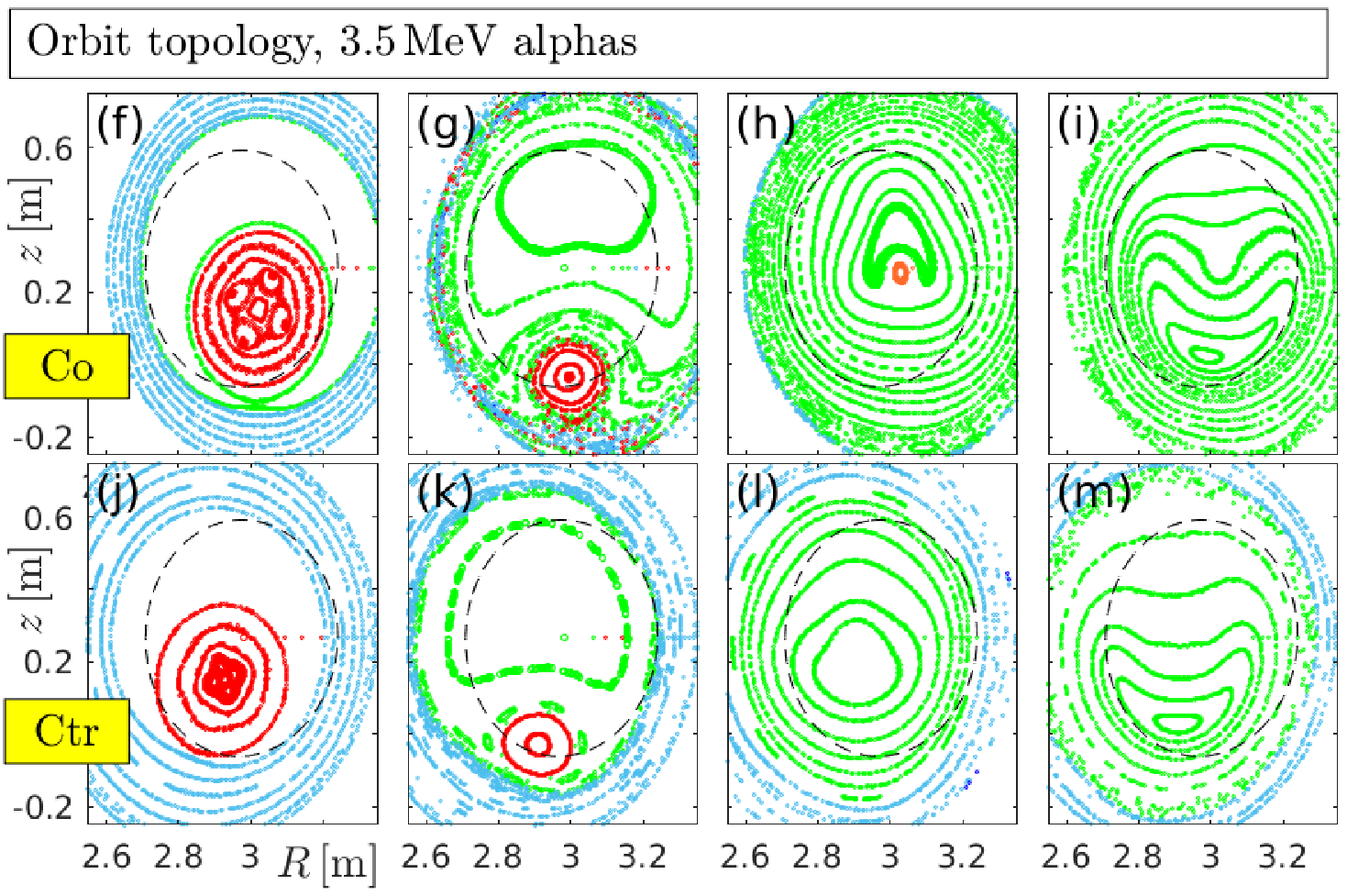}\vspace{-1.1cm} \\
\includegraphics[width=0.45\textwidth]{\figures/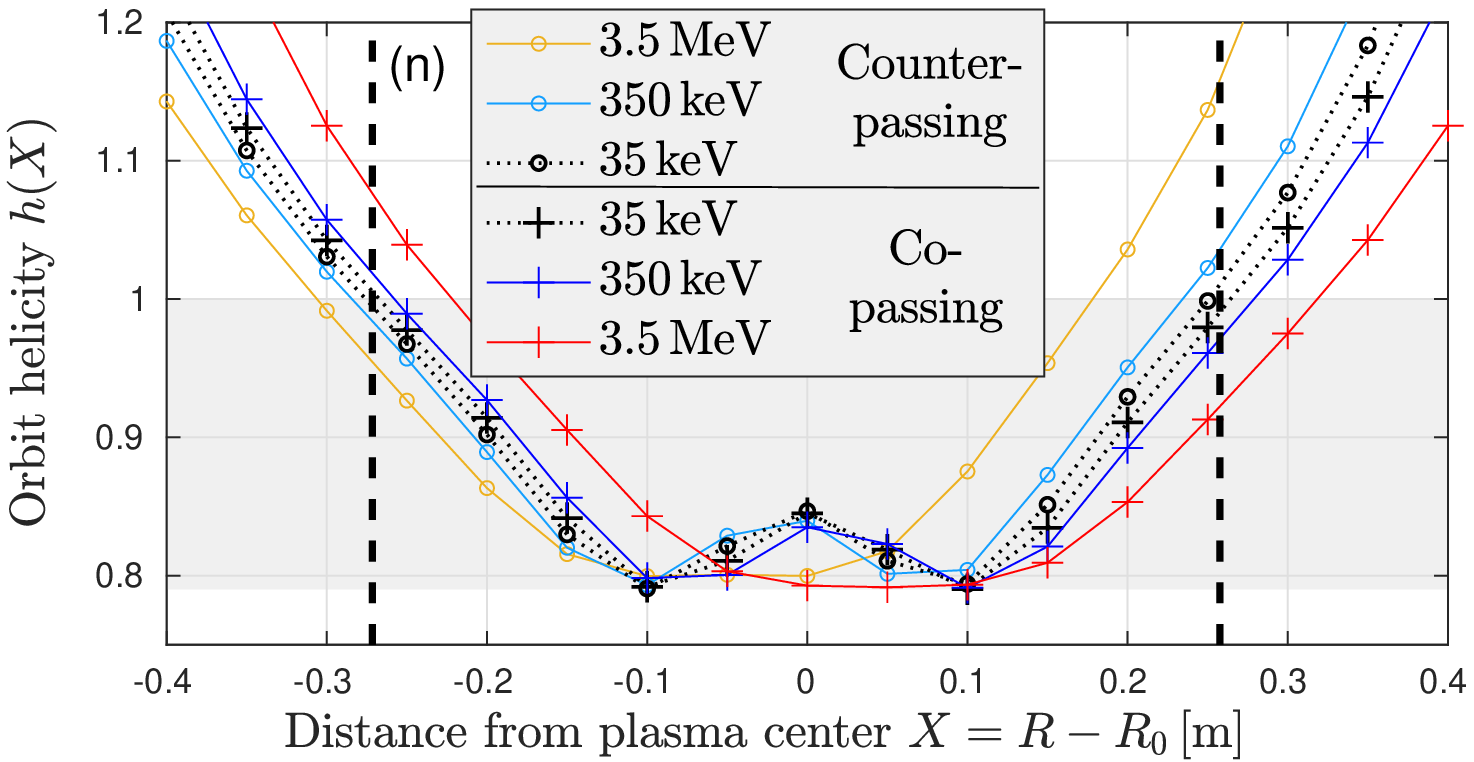}
\caption{Overview of the sawtooth crash dynamics in the case with non-monotonic $q$ profile and central value $q_0 = 0.87$. Arranged as Fig.~\protect\ref{fig:s11_scan2_poin}.}
\label{fig:s12_scan4_poin}%
\end{figure}

\subsection*{7. Safety factor scan}

The physical picture proposed in the main article implies that the selective confinement of fast alphas is sensitive with respect to the profile of the field helicity (safety factor) $q(\overline{r})$. In order to highlight this effect, we present in this section preliminary results obtained from simulations with $q$ profiles that have a central value $q_0$ lower than our default case with $q_0 = 0.98$. The two cases considered are shown in Fig.~\ref{fig:s02_prof_qsafe}: a monotonic $q$ profile with $q_0 = 0.95$, and a non-monotonic $q$ profile with $q_0 = 0.87$ and an off-axis minimum $q_{\rm min} = 0.79$ at $\overline{r} \approx 0.1\,{\rm m}$. Since the $q = 1$ radius is larger in these cases, we have increased the size of the simulation domain from $25\%$ to $50\%$ of the poloidal magnetic flux space.

The evolution of the MHD fluctuation energy in these two cases is shown in Fig.~\ref{fig:s10_evol_enr-q_scan24}(a,c). One can see that the $n=1$ harmonic dominates in both simulations. Comparison with Fig.~\ref{fig:s03_evol_enr_alpha}(a,b) shows that the growth rate of the internal kink in Fig.~\ref{fig:s10_evol_enr-q_scan24} has increased with decreasing $q_0$, which --- to be precise --- is not primarily due to the value of $q_0$ itself but due to the steeper gradient that these $q(\overline{r})$ profiles have at the $q = 1$ radius, and which translates to a steeper gradient in the plasma current. This results in a higher $\ExB$ velocity and a shorter crash time as one can see in Figs.~\ref{fig:s11_scan2_poin}(a) and \ref{fig:s12_scan4_poin}(a): the vertical displacement velocity $v_{z0}$ at the magnetic axis reaches nearly $10\,{\rm km/s}$ in the case with $q_0 = 0.95$, and exceeds $20\,{\rm km/s}$ in the case with $q_0 = 0.87$. The implications for fast alpha confinement will be discussed shortly.

Before proceeding, we would like to insert a note of caution. Although low central values of the field helicity like $q_0 \sim 0.87$ in our third model scenario are consistent with some experimental observations, this parameter alone does not determine the sawtooth crash dynamics. For instance, the crash can be expected to proceed more slowly when the $q$ profile is initialized with a flat  `shoulder' around the $q = 1$ radius \cite{Soltwitsch87, Kolesnichenko92}. This is one of the reasons why the present safety factor scan has only a preliminary character: the exact shape of the profile matters, so the cases we have simulated and discuss here should be viewed as mere random examples.

Panels (b) and (d) of Fig.~\ref{fig:s10_evol_enr-q_scan24} show the evolution of $q$ profiles in the cases with $q_0 = 0.95$ and $0.87$. These plots show how the central portion of the $q$ profile tends to rise above unity during the crash phase and then oscillates within a few percent around $q \sim 1$ until the $\ExB$ flow has decayed to an insignificant level (and sources begin to restore the initial profile). These perturbed profiles must also be interpreted with care because we have simply averaged the field helicity over the {\it unperturbed} flux surfaces (each identified by a unique value of $\overline{r}$) of the initial MHD equilibrium. Ideally, the perturbed helicity profiles should be computed in accordance with the perturbed magnetic surfaces in order to be meaningful. Since this was not done here, the plots in Fig.~\ref{fig:s10_evol_enr-q_scan24}(b,d) are meant to show only the overall trend.

The actual evolution of the magnetic topology can be seen in Figs.~\ref{fig:s11_scan2_poin}(b-e) and \ref{fig:s12_scan4_poin}(b-e). Both cases exhibit a Kadomtsev-type crash followed by Wesson-type overshoots of the $\ExB$ flow. At later times, panel (a) in both figures shows a temporary reversal of the flow direction at the magnetic axis, which seems to be a signature of increasingly turbulent motion (perhaps the kink's return flows become Kelvin-Helmholtz unstable). Note that the volume integrated MHD fluctuation energy in Fig.~\ref{fig:s10_evol_enr-q_scan24} is still large at that time.

The Poincar\'{e} plots of the orbit topology show that there are still readily visible differences between co- and counter-passing $3.5\,{\rm MeV}$ alphas in the case with $q_0 = 0.95$ in Fig.~\ref{fig:s11_scan2_poin}, while the overall structures begin to look similar in the case with $q_0 = 0.87$ in Fig.~\ref{fig:s12_scan4_poin}. This, of course, can be expected from the fact that the magnitude of the magnetic drifts remains more or less the same, while the $q = 1$ radius increases in our two model profiles.

\begin{figure}
[tbp]
\centering\vspace{-0.25cm}
\includegraphics[width=0.45\textwidth]{\figures/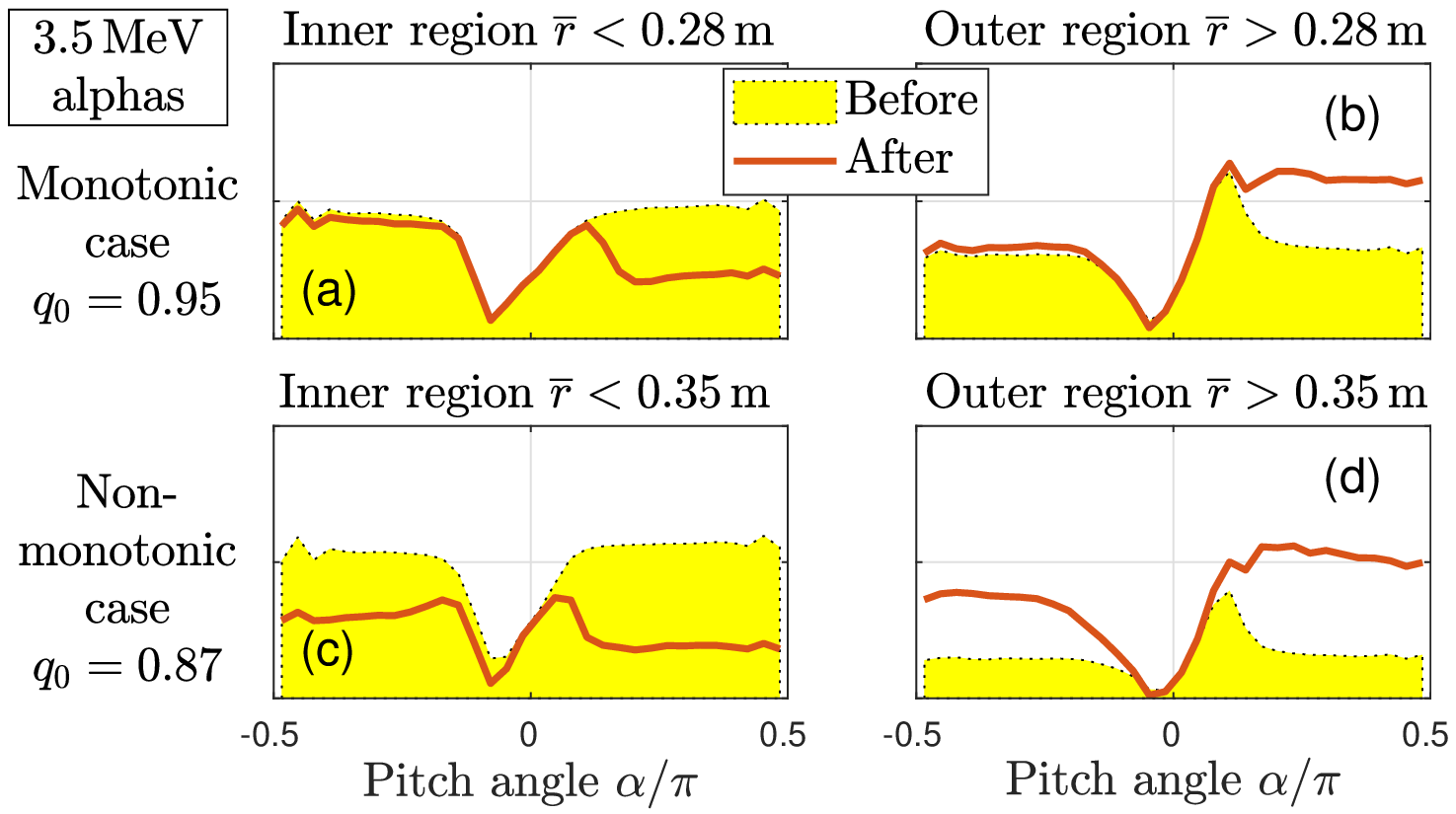}\vspace{-0.15cm}
\caption{Pre- and post-crash distributions in pitch angle $\alpha = \sin^{-1}(v_\parallel/v)$ for $3.5\,{\rm MeV}$ alphas in the cases with $q_0 = 0.95$, $\overline{r}_1 = 0.24\,{\rm m}$ (a,b) and $q_0 = 0.87$, $\overline{r}_1 = 0.3\,{\rm m}$ (c,d). The boundary between integration regions (I) and (II) is chosen to be somewhat larger than the respective $q = 1$ radius $\overline{r}_1$. Arranged as Fig.~\protect\ref{fig:05_pitch}.}
\label{fig:s13_scan24_pitch_3500keV}%
\end{figure}

In contrast to our default case in Fig.~\ref{fig:07_helicity}, the orbit pitch profiles in panel (n) of Figs.~\ref{fig:s11_scan2_poin} and \ref{fig:s12_scan4_poin} all lie below unity in the plasma center, so there are $h = 1$ resonances and hence reconnection occurs in the orbit topology of both co- and counter passing $3.5\,{\rm MeV}$ alphas. However, in the case with $q_0 = 0.95$, the $h = 1$ resonance lies almost entirely within the $q = 1$ radius $\overline{r}_1 \approx 0.24\,{\rm m}$, so that these particles tend to be mixed only inside that region. This can be verified in Fig.~\ref{fig:s13_scan24_pitch_3500keV}(a,b), which shows that effectively no transport occurs at negative pitch angles $\alpha < 0$ across the radius $\overline{r} = 0.28\,{\rm m}$ that we have chosen as a boundary between the inner region (I) and the outer region (II). The situation is different in the case with $q_0 = 0.87$, where a part of the $h = 1$ resonance lies outside the $q = 1$ radius $\overline{r}_1 \approx 0.3\,{\rm m}$ and where the mixing radius extends far beyond the initial $q = 1$ radius due to the larger amount of reconnected magnetic flux (e.g., see Fig.~8.12 of \cite{WhiteTokBook3}). For this case, Fig.~\ref{fig:s13_scan24_pitch_3500keV}(c,d) shows that there is significant transport of both co- and counter-passing $3.4\,{\rm MeV}$ alphas across the chosen boundary $\overline{r} = 0.35\,{\rm m}$ between region (I) and (II).

Another notable and interesting difference between the two cases that can be seen in the Poincar\'{e} plots in Figs.~\ref{fig:s11_scan2_poin} and \ref{fig:s12_scan4_poin} is that the field and orbit topology in the monotonic case with $q_0 = 0.95$ still consists of good Kolmogorov-Arnold-Moser (KAM) surfaces on global scales, while the non-monotonic case with $q_0 = 0.87$ exhibits signatures of chaos near the boundary of the $q \sim 1$ (Fig.~\ref{fig:s12_scan4_poin}(e)) and $h \sim 1$ domain (Fig.~\ref{fig:s12_scan4_poin}(g,k) and onward). From the theory of nonlinear oscillators, it is known that such behavior is an indication of the existence of multiple resonances that have different helicities and coexist in the same domain (= nonlinear resonance overlap).

While the consequences of resonance overlaps are understood, the reason for why multiple resonances coexist in our non-monotonic case with $q_0 = 0.87$ but not in the other cases remains to be clarified. We note that, during the early stages of the sawtooth crash, the non-monotonic case has a resonance with $h = 4/5 > q_{\rm min} = 0.79$. Having a poloidal peridicity of $4$, this resonance is likely to be the reason for the island chain that can be seen near the orbit axis in panels (f) and (j) of Fig.~\ref{fig:s12_scan4_poin}. Similar structures were reported in \cite{Kolesnichenko98}. Indeed, at the time of this snapshot ($t - t_0 \approx 0.07\,{\rm ms}$, $\hat{t} \approx 500$) the minimum of the $q$ profile in Fig.~\ref{fig:s10_evol_enr-q_scan24}(d) is still below $0.8$. Shortly after (not shown) we observe an island chain with poloidal periodicity of $5$ in the co-passing orbits, which is indicative of an $h = 5/6 = 0.8\overline{3}$ resonance in close proximity of the reconnection layer. The topological distortions associated with these islands may also couple to the above-mentioned modulations of MHD flows (possibly leading to MHD turbulence). One phenomenon may be the trigger of the other, or they may both be manifestations of one `secondary nonlinear instability'. Whatever is happening, it leads here to the formation of a chaotic belt, especially in the case of co-passing particles in Fig.~\ref{fig:s12_scan4_poin}(g).

\begin{figure}
[tbp]
\centering
\includegraphics[width=0.45\textwidth]{\figures/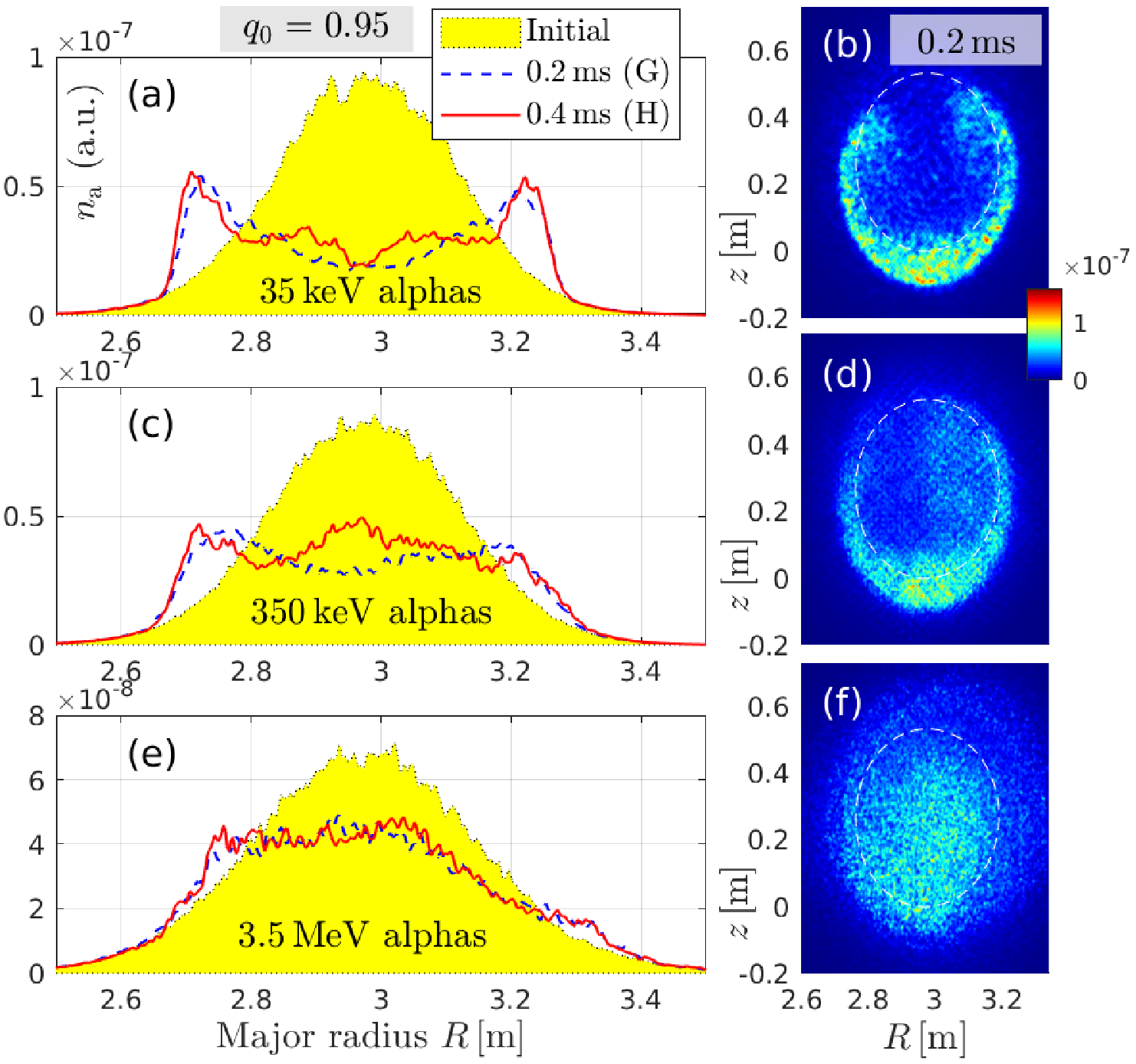}
\caption{Spatial transport of alpha particles with kinetic energies $K = 35\,{\rm keV}$ (a,b), $350\,{\rm keV}$ (c,d), $3.5\,{\rm MeV}$ (e,f) during the sawtooth crash in Fig.~\protect\ref{fig:s11_scan2_poin} in the case with $q_0 = 0.95$. The snapshot times (G) and (H) were indicated in Fig.~\protect\ref{fig:s10_evol_enr-q_scan24}(a). Arranged as Fig.~\protect\ref{fig:04_mega_kin_alpha}, but showing a larger spatial domain. These data were obtained in simulations of a reduced domain covering the inner 50\% of magnetic flux space.}
\label{fig:s14_scan2_kin_alpha}%
\end{figure}

\begin{figure}
[tbp]
\centering
\includegraphics[width=0.45\textwidth]{\figures/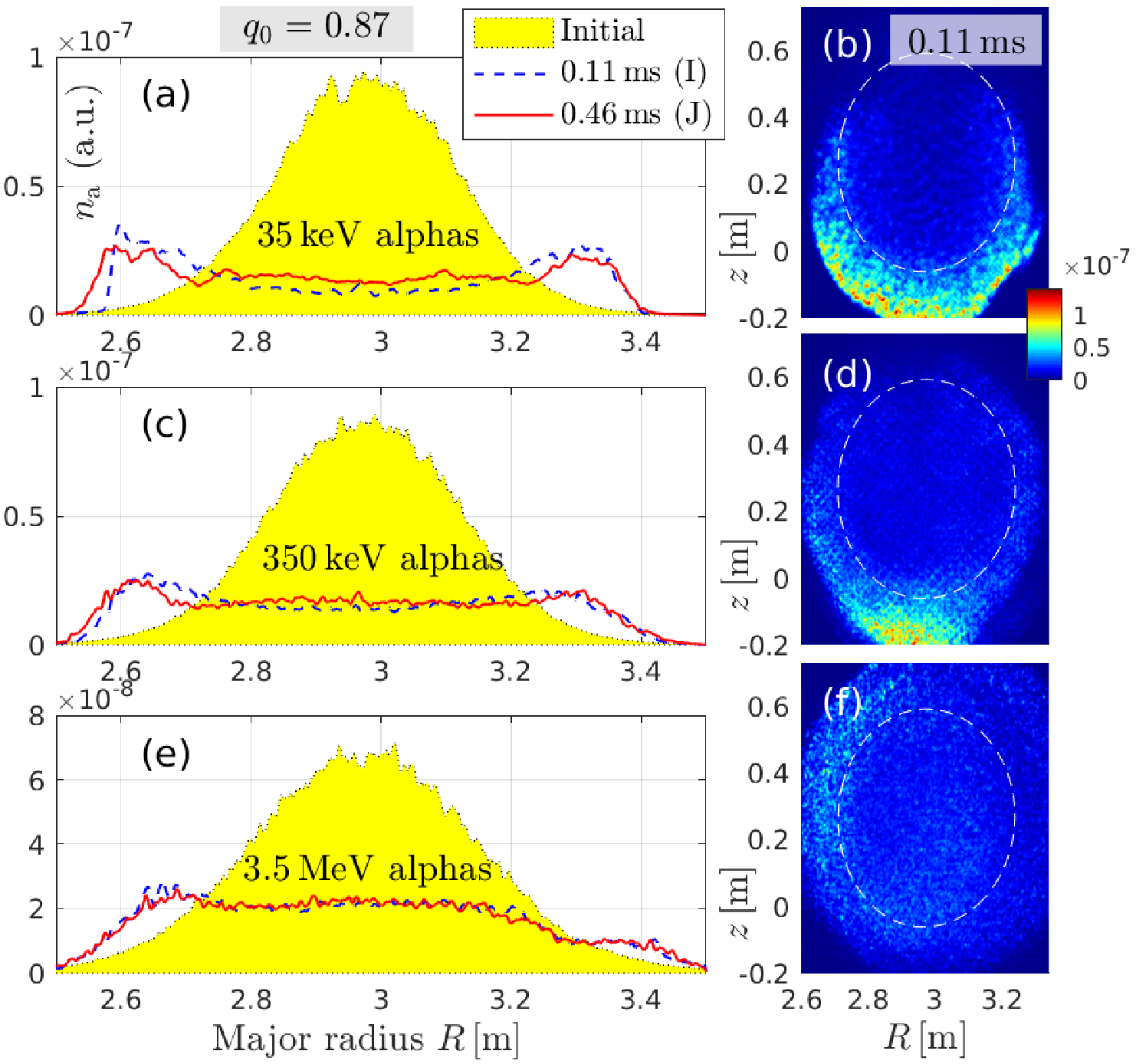}
\caption{Spatial transport of alpha particles with kinetic energies $K = 35\,{\rm keV}$ (a,b), $350\,{\rm keV}$ (c,d), $3.5\,{\rm MeV}$ (e,f) during the sawtooth crash in Fig.~\protect\ref{fig:s12_scan4_poin} in the case with $q_0 = 0.87$. The snapshot times (I) and (J) were indicated in Fig.~\protect\ref{fig:s10_evol_enr-q_scan24}(c). Arranged as Fig.~\protect\ref{fig:04_mega_kin_alpha}, but showing a larger spatial domain. These data were obtained in simulations of a reduced domain covering the inner 50\% of magnetic flux space.}
\label{fig:s15_scan4_kin_alpha}%
\end{figure}

Finally, we inspect the spatial transport of monoenergetic alpha particles with $K = 35\,{\rm keV}$, $350\,{\rm keV}$ and $3.5\,{\rm MeV}$, results for which are summarized in Figs.~\ref{fig:s14_scan2_kin_alpha} and \ref{fig:s15_scan4_kin_alpha}. In both cases, fast alphas with $3.5\,{\rm MeV}$ are still better confined than slow alphas with $35\,{\rm keV}$. However, in comparison with our default case with $q_0 = 0.98$ in Fig.~\ref{fig:04_mega_kin_alpha}(e), the reduction of the fast alpha density in the plasma center has become significant: a drop by about $30\%$ and $60\%$ is seen in Fig.~\ref{fig:s14_scan2_kin_alpha}(e) and \ref{fig:s15_scan4_kin_alpha}(e), respectively. Obviously, such kinds of sawtooth crashes are less attractive for the control of helium ash than our default case in Fig.~\ref{fig:04_mega_kin_alpha}.

The trend is consistent with the underlying physical picture that we have described in the main article. We have already seen in Figs.~\ref{fig:s11_scan2_poin}(n) and \ref{fig:s12_scan4_poin}(n) that the $h = 1$ resonances exist also for counter-passing particles in these cases, so that their orbit topology is subject to reconnection. In other words, by reducing $q_0$, we have lost much of the orbit topology's sensitivity with respect to magnetic drifts that was shown in box (ii) of Fig.~\ref{fig:06_synergy}.

More detailed analysis is required to evaluate the ratio $\tau_{2\pi}/\tau_{\rm crash}$ of resonance detuning and sawtooth crash times, whose role was illustrated in boxes (iii) and (iv) of Fig.~\ref{fig:06_synergy}. A first rough inspection indicates that this part of the synergism is also broken. We have already seen in Figs.~\ref{fig:s11_scan2_poin} and \ref{fig:s12_scan4_poin} that the crash times $\tau_{\rm crash}$ have become shorter by about a factor 2--4. This may be partly compensated by the larger pre-crash values of the parameter $|1 - h|$, which would imply a correspondingly shorter detuning time $\tau_{2\pi}$. However, the perturbed $q$ profiles in Fig.~\ref{fig:s10_evol_enr-q_scan24}(b,d) approach unity at the time when the $\ExB$ velocity peaks, so the detuning time $\tau_{2\pi}$ may in fact be as long as in the $q_0 = 0.98$ case, implying that even fast alphas are unable to detach from the kink due to the shorter crash time $\tau_{\rm crash}$.

Evidence for this can be seen in the redistribution of $350\,{\rm keV}$ alphas in Fig.~\ref{fig:s14_scan2_kin_alpha}(d) for $q_0 = 0.95$, which develop the same horse-shoe-like structure as the $35\,{\rm keV}$ alphas in Fig.~\ref{fig:04_mega_kin_alpha}(b) for $q_0 = 0.98$. This implies that the energy threshold beyond which alpha particles can decouple from the kink has increased substantially, possibly even above the energy of newly-born $3.5\,{\rm MeV}$ alphas.

In the case with $q_0 = 0.87$, off-axis humps can be seen in the post-crash density profiles even for $3.5\,{\rm MeV}$ alphas, which implies that their redistribution has become akin to that of an MHD fluid. A theory for this regime was proposed by Kolesnichenko {\it et al}.\ \cite{Kolesnichenko92}, and this has been suggested as an explanation for observed redistribution of alpha particles in the intermediate energy range $150...600\,{\rm keV}$ in TFTR experiments \cite{Stratton96}. Although the measurements in the plasma center suffered from relatively large uncertainties \cite{McKee97}, our simulation results in Figs.~\ref{fig:s14_scan2_kin_alpha} and \ref{fig:s15_scan4_kin_alpha} support this possibility in principle.

Note that the $3.5\,{\rm MeV}$ alpha particle density profile in Fig.~\ref{fig:s15_scan4_kin_alpha}(e) has developed a hump on the high-field side ($R \approx 2.7\,{\rm m}$) and a shoulder on the low-field side ($R \approx 3.3\,{\rm m}$). This is most likely a manifestation of the different redistribution of co- and counter-passing alpha profiles, showing that magnetic drifts have clearly visible effects even in the case with low $q_0 = 0.87$.

\end{document}